\documentclass[a4paper]{article}
\usepackage{graphicx}

\newcommand{\rg}{r_{\rm g}}
\newcommand{\mbh}{M_{\bullet}}
\newcommand{\sun}{\odot}
\newcommand{\loc}{{\rm l}}
\renewcommand{\vec}[1]{\mbox{\protect\boldmath$#1$}}

\begin{document}

\title{An {\fontfamily{phv}\fontshape{sc}\selectfont xspec} model to explore
spectral features\\ from black-hole sources\footnote{In proceedings of
the Workshop {\it{}Processes in the vicinity of black holes and neutron
stars} (Silesian University, Opava), in press.}}
\author{M.~Dov\v{c}iak,\footnote{Astronomical Institute, Academy of Sciences,
 Prague, Czech Republic}~\footnote{Faculty of Mathematics and Physics,
 Charles University Prague, Czech Republic}~
V.~Karas,\footnotemark[2]~\footnotemark[3]~
A.~Martocchia,\footnote{Observatoire Astronomique, Strasbourg, France}~
G.~Matt,\footnote{Dipartimento di Fisica, Universit\`a degli Studi ``Roma Tre'',
 I-00146~Roma, Italy}~
 and
T.~Yaqoob\footnote{Department of Physics and Astronomy,
 Johns Hopkins University, Baltimore, MD~21218}~~\footnote{Laboratory for High
 Energy Astrophysics, NASA/Goddard Space Flight Center, Greenbelt, MD~20771}}
\maketitle
\begin{abstract}
We report on a new general relativistic computational model enhancing,
in various respects, the capability of presently available tools for
fitting spectra of X-ray sources. The new model is intended for spectral
analysis of black-hole accretion discs. Our approach is flexible enough
to allow easy modifications of intrinsic emissivity profiles. Axial
symmetry is not assumed, although it can be imposed in order to reduce
computational cost of data fitting. The main current application of our
code is within the {\sc xspec} data-fitting package, however, its
applicability goes beyond that: the code can be compiled in a
stand-alone mode, capable of examining time-variable spectral features
and doing polarimetry of sources in the strong-gravity regime. Basic
features of our approach are described in a separate paper (Dov\v{c}iak,
Karas \& Yaqoob \cite{dovciak04}). Here we illustrate some of its
applications in more detail. We concentrate ourselves on various aspects
of line emission and Compton reflection, including the current
implementation of the lamp-post model as an example of a more
complicated form of intrinsic emissivity.
\end{abstract}

\section{Introduction}
Regions of strong gravitational field are most usually explored via
X-ray spectroscopy, because very hot X-ray emitting material is commonly
believed to be present in regions near a neutron star surface or
a black-hole horizon. Accretion plays crucial role in the process of
energy liberation and mass accumulation that takes place in this kind of
objects \cite{kato98,krolik99}. In particular, disc-type accretion
represents an important mode
which is realized under suitable circumstances, given by the global
geometrical arrangement and local microphysics of the fluid medium. A
central compact body, undetectable via its own radiation, resides in
galactic nuclei where it is surrounded by a rather dense population of
stars and gaseous environment. Photons emerging from the accretion disc
and its corona are influenced by gravity of the black hole, so they
bear various imprints of the gravitational field structure. This
concerns especially X-rays originating very near the core and, for this
reason, spectral analysis with X-ray satellites is particularly relevant
for astronomical study of strong gravitational fields around black
holes. For a general discussion, see review articles
\cite{fabian00,reynolds03} and further references cited therein.

In this paper we describe a newly developed computational model aimed
for the spectral analysis of line profiles and continuum originating in
a geometrically thin, planar accretion disc near a rotating (Kerr) black
hole. Such analysis has been routinely performed via {\sc xspec} package
\cite{arnaud96}, which performs deconvolution of observed spectra for the
effective area and energy redistribution of the detector. Previously,
several routines were developed and linked with this package in order to
fit data to a specific model of a black-hole accretion disc 
\cite{laor_1991,martocchia00}. However, a substantially
improved variant of the computational approach has been desirable
because previous tools have various limitations that may be critical for
analyses of present-day and forthcoming high-resolution data.

We describe the layout and usage of the new code and we show some
examples and comparisons between new model components. Our present
contribution provides information complementary to the basic description
which can be found in ref.\ \cite{dovciak04}. We suggest the
reader to consult that paper as well as further details in Thesis
\cite{dovciak04a}, as they give more examples and
citations to previous works. Here we concentrate our attention to
technical issues of the code structure and its performance when
computing and fitting spectra. Different perspectives and
applications of general relativistic computations for black-hole
accretion discs have been considered by various authors. In particular,
it is very useful to consult recent papers of Gierli\'{n}ski,
Maciolek-Nied\'{z}wiecki  \& Ebisawa \cite{gierlinski01} and Schnittman
\& Bertschinger \cite{schnittman04}. Very recently, a new independent
code has been developed by Beckwith \& Done \cite{beckwith04}.
Their approach also allows to study accretion disc spectra including
strong gravity  effects of a Kerr black hole. This is also one of the
applications of our code, and  so relatively accurate comparisons
between both tools are possible. We performed several such comparisons
and found  a very good agreement in the shape of predicted line
profiles. 

Given a limited space for this contribution, we
cannot describe all aspects of the new code: capability of the code
with respect to timing and polarimetry are discussed elsewhere.
Nonetheless, it may be good to bear in mind that such capability
has been already implemented and tested, taking into account 
all strong-gravity effects on time-delays and the Stokes parameters. 

\begin{figure*}[tbh]
{\small
\begin{tabular}{p{0.32\textwidth}p{0.32\textwidth}p{0.32\textwidth}}
\includegraphics[width=0.3\textwidth]{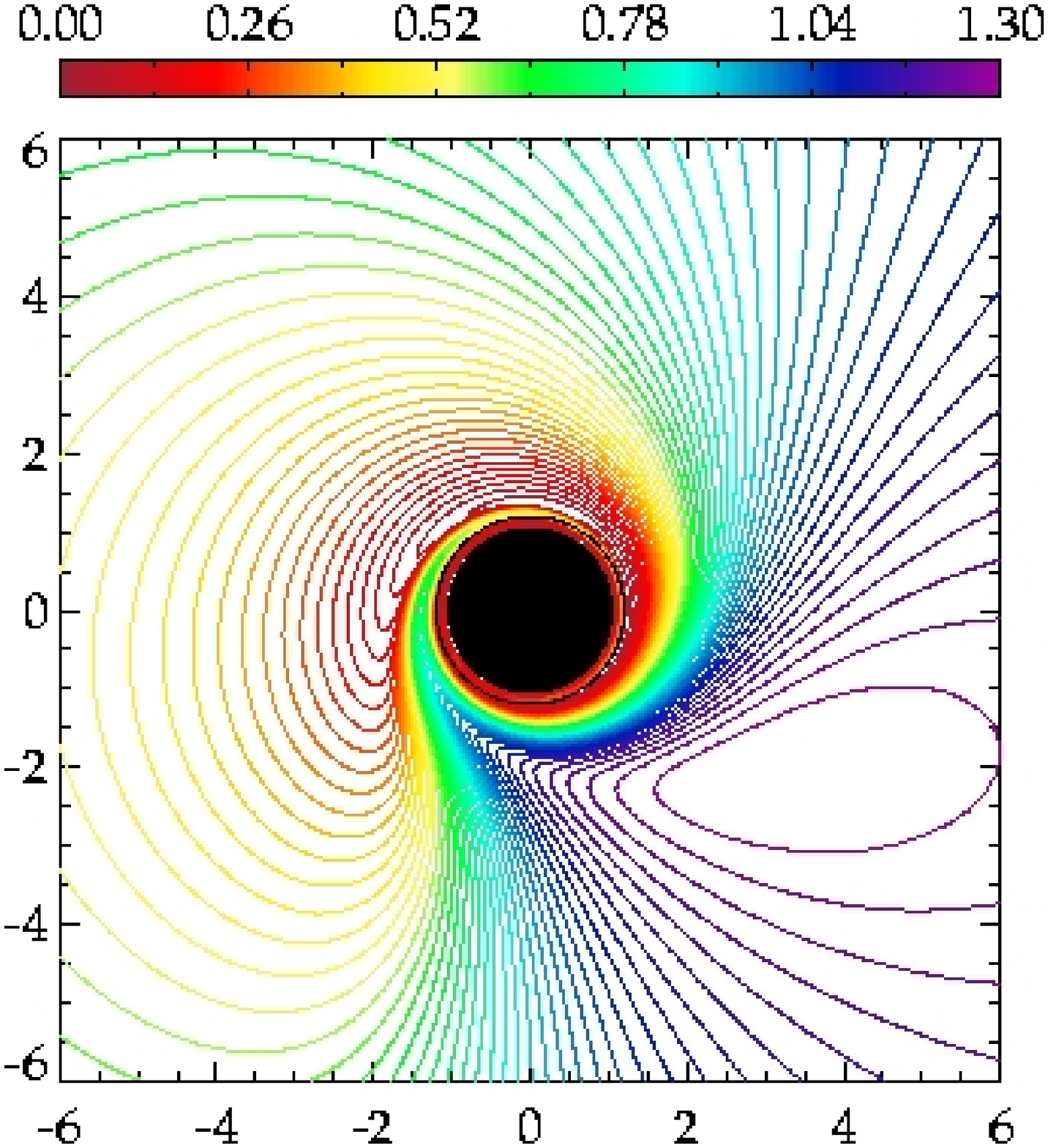}
&
\includegraphics[width=0.3\textwidth]{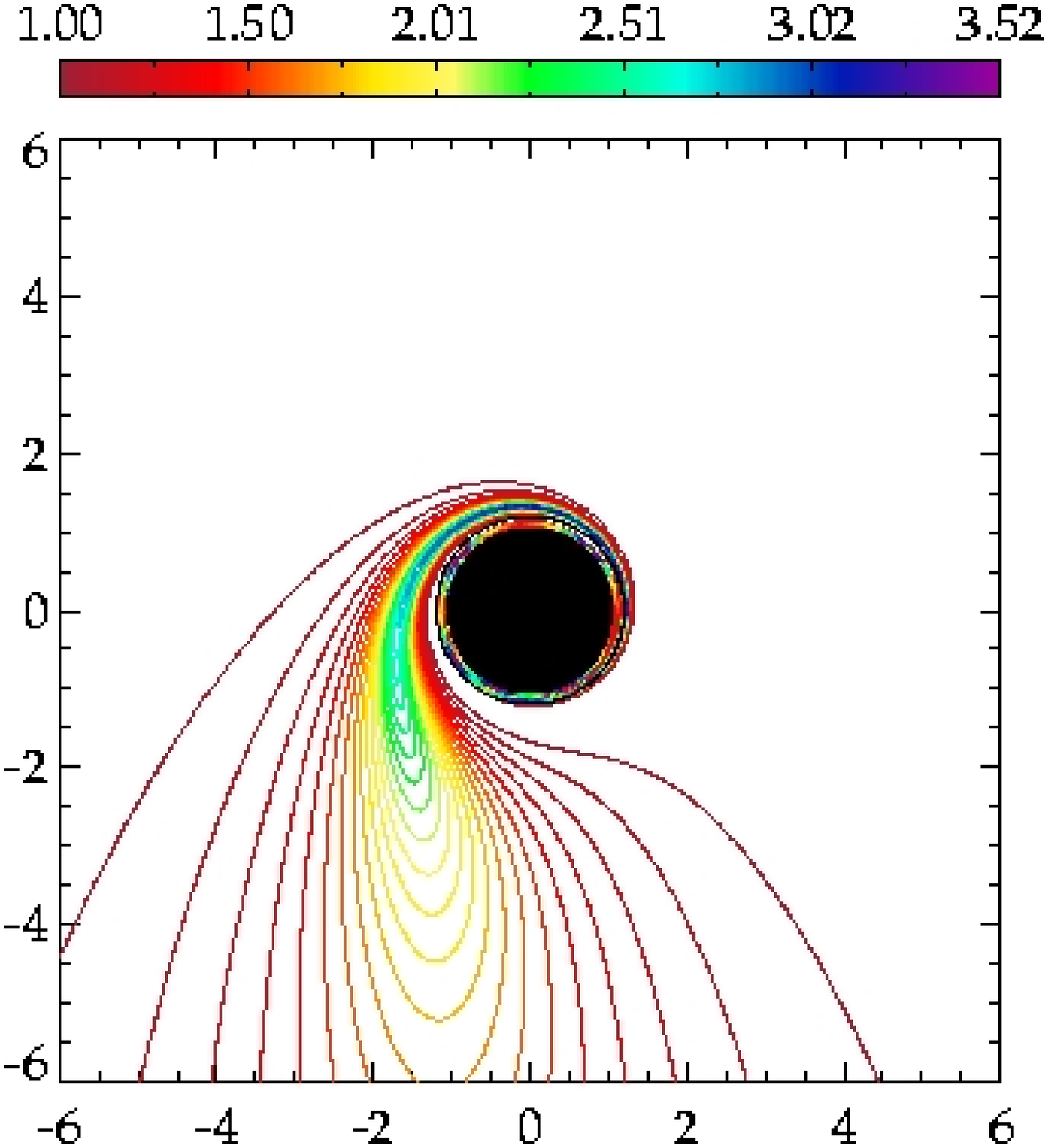}
&
\includegraphics[width=0.3\textwidth]{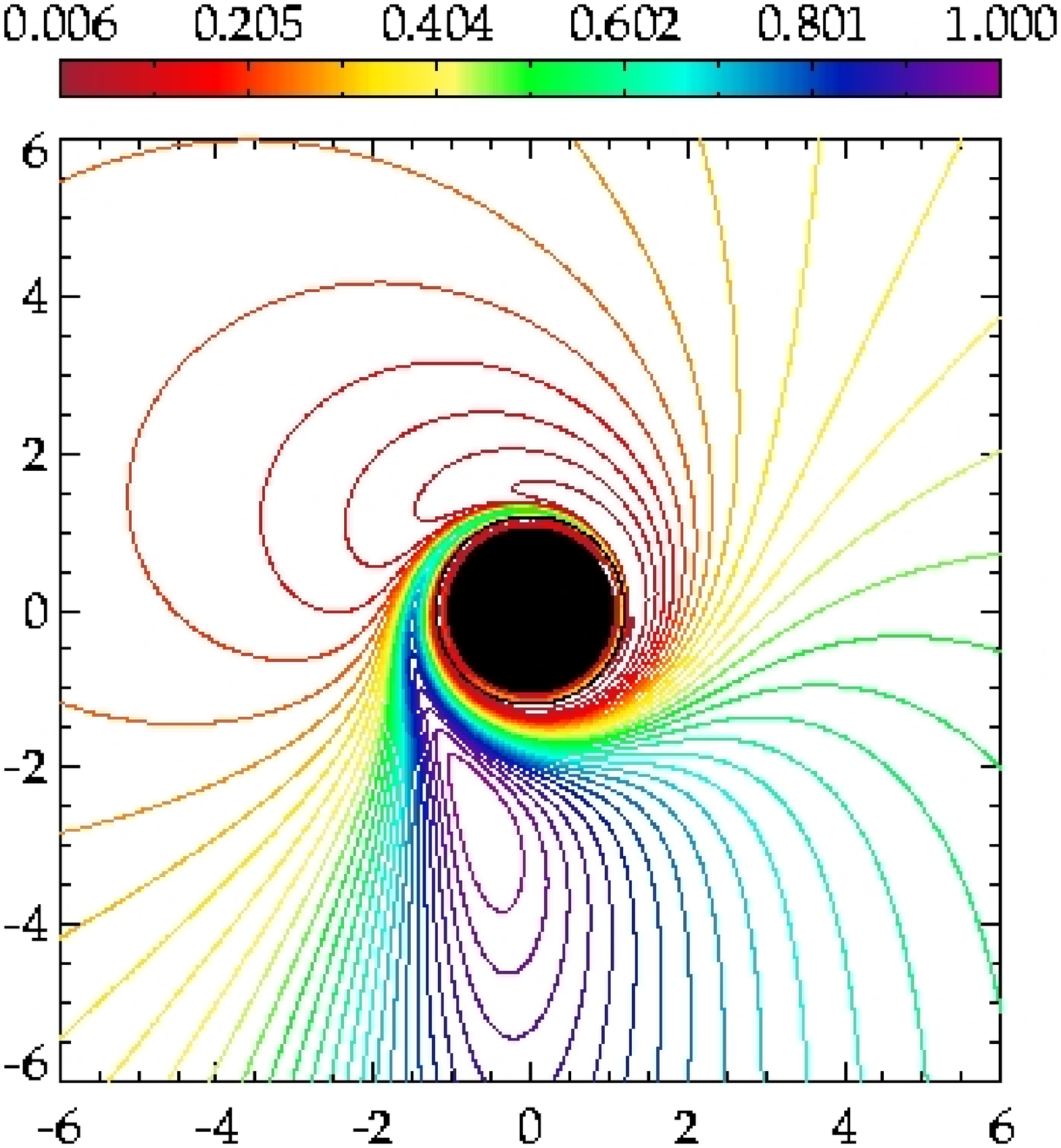}
\\
a) $g$-factor & b) lensing &
c) \parbox[t]{3.6cm}{cosine of the angle of emission} \\[5mm]
\includegraphics[width=0.3\textwidth]{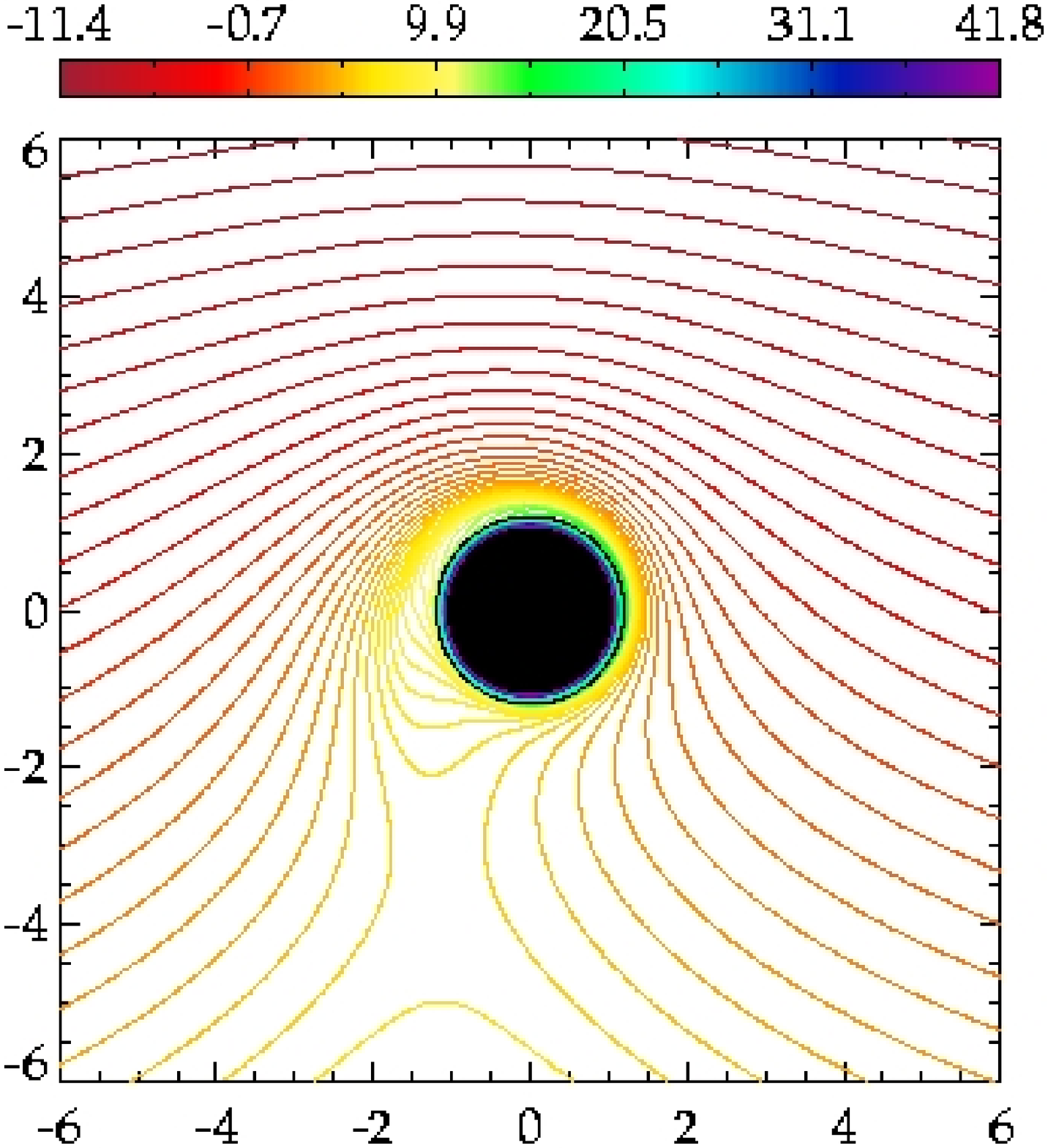}
&
\includegraphics[width=0.3\textwidth]{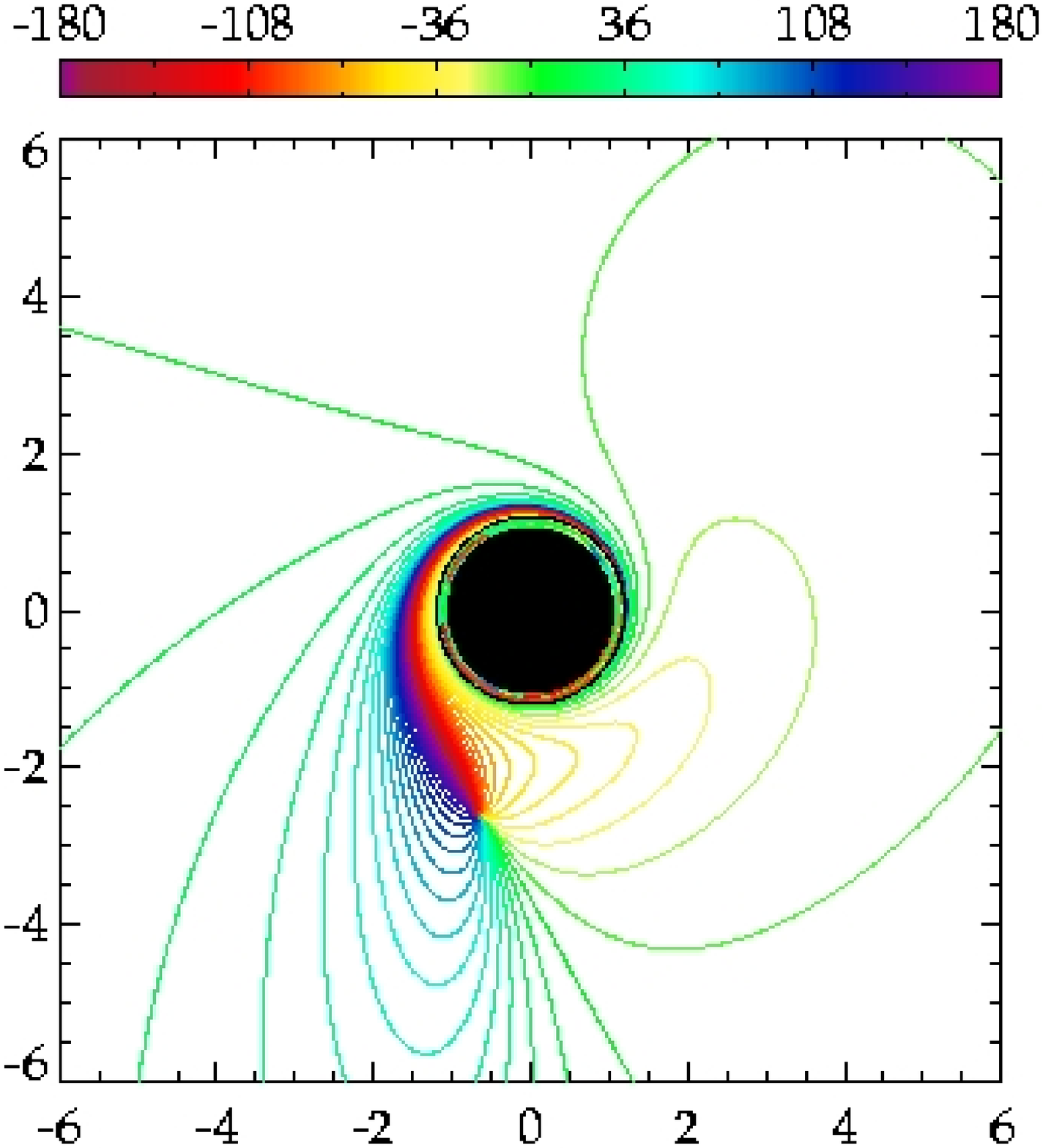}
&
\includegraphics[width=0.3\textwidth]{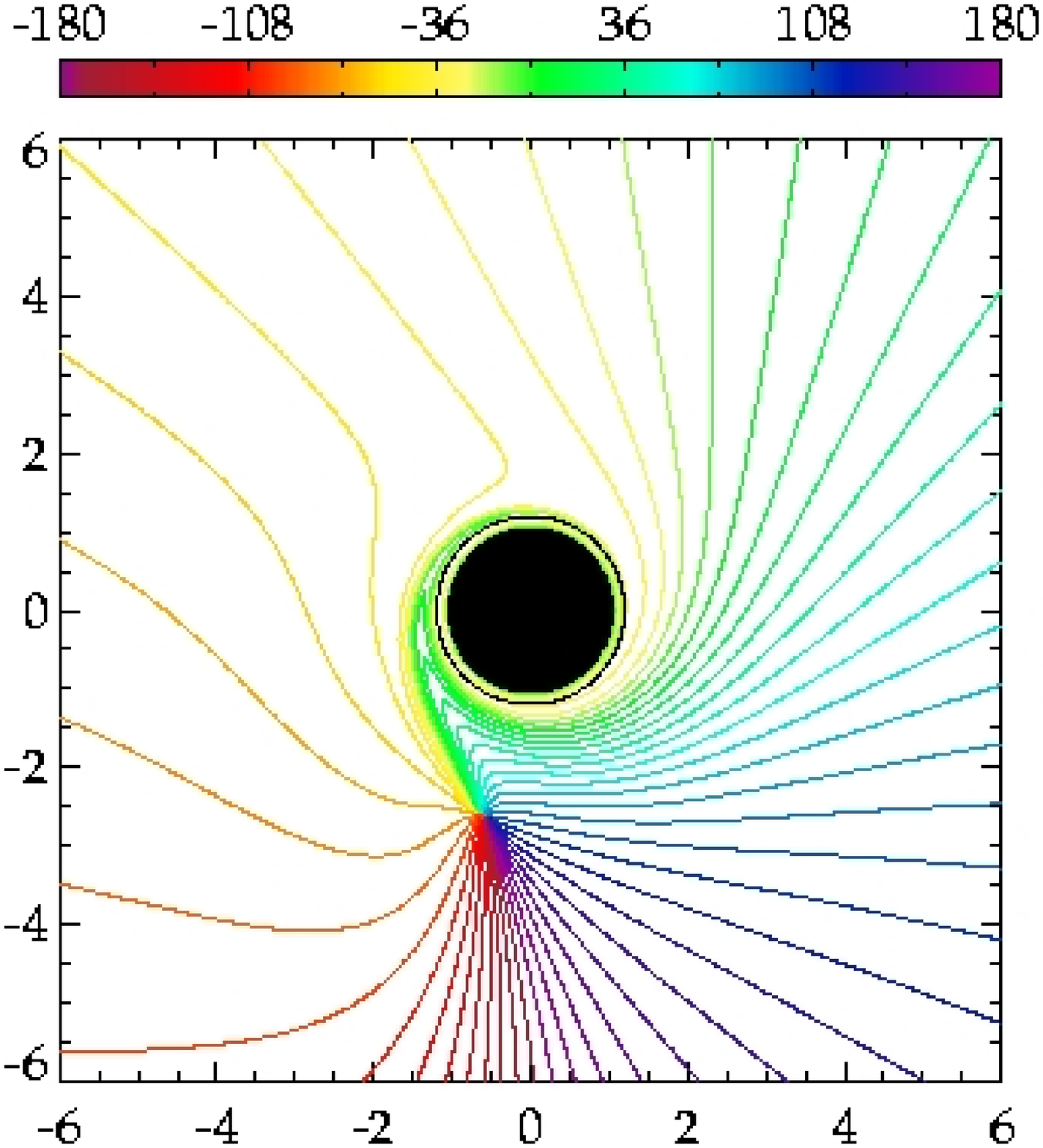}
\\
d) relative time delay & e) \parbox[t]{3.6cm}{change of the angle of polarization} &
f) \parbox[t]{3.6cm}{azimuthal emission angle}\\
\end{tabular}}
\caption{Six transfer functions are shown in the equatorial plane
of a Kerr black hole with $a\doteq0.9987$ (horizon $r_{\rm h}=1.05$). 
The black hole rotates counter-clockwise. The observer is located 
upwards at the inclination $\theta_{\rm o}=70^\circ$. 
The values of transfer functions are encoded by a colour scale, as
indicated above each graph. For mathematical formulae defining these
functions see eqs.~(\ref{gfac}), (\ref{lensing})--(\ref{polar}) 
and (\ref{azim_angle}) in 
Appendix~\ref{appendix1}.}
\label{ftransf}
\end{figure*}

\section{Transfer functions}
\label{transfer_functions}
We concentrate on geometrically thin and optically thick accretion discs
and we point out that general relativity effects can play a role if the
configuration is sufficiently dense in a limited region, typically a few
tens of gravitational radii. Nevertheless, our computational domain 
extends up to about $\approx10^3$ gravitational radii in a non-uniform
spatial grid.

In order to calculate the final spectrum that an observer at infinity
measures when local emission from the accretion disc is given, one must
first specify the intrinsic emissivity in frame co-moving with the disc
medium and then perform transfer of photons to a distant observer. Here
we concentrate on the latter part of this task.

Six functions need to be computed across the source: (i)~energy shift
affecting the photons (i.e.\ gravitational and Doppler $g$-factor,
needed to account for spectral redistribution), (ii)~gravitational
lensing (for the evaluation of radiation flux or count rate),
(iii)~direction of emission with respect to the disc normal (for the
limb darkening effect), (iv)~relative time-delay of the light signal
(i.e.\ the mutual delay between photons arriving from different regions
of the source, needed for the proper account of the light-time effect 
in timing analysis), (v)~change of the polarization angle due to
photon propagation in the gravitational field (for
polarimetry), and (vi)~azimuthal direction of emission (also for
polarimetry). While the first three quantities are always necessary, the
time-delay factor is required only when local emission is not stationary
(e.g.\ the case of orbiting spots) and the change of the polarization angle
with the azimuthal direction of emission are essential for calculating
the overall degree and angle of polarization, as observed at infinity. In
the adopted approximation of geometrical optics, light rays follow null
geodesics (in curved space-time) and spectral computations are reduced to
a text-book problem \cite{chandra92} which, however, may be rather
demanding computationally on practical level of data fitting. Useful
form of light-ray equations and further references can be found in
various papers \cite{dovciak04,fanton97,rauch94}. We summarize the
basic equations in Appendix~\ref{appendix1}.

In order for our new model to be fast and practical, we pre-calculated
the transfer functions for $21$ values of the angular momentum of the black
hole and $20$ values of the inclination angle of the observer. The choice of
the grid appears sufficiently fine to ensure high accuracy. We have stored
the transfer functions in the form of tables -- i.e.\ as binary extensions 
of a FITS file. For a technical description of the files layout see
Appendix~\ref{appendix3a}. Values of the transfer functions are
interpolated when integrating the spectrum for a given angular momentum
of the black hole and inclination angle of the observer.

The graphical representation of the tables is shown in Figure~\ref{ftransf}.
Six frames of contour plots correspond to individual transfer functions,
which are necessary in computations. This figure captures equatorial
plane for given values of $a$ and $\theta_{\rm o}$. The radius extends up 
to $r=10^3\rg$ in the tables, but here we show only the central region,
$r\leq6\rg$, where relativistic effects are most
prominent.\footnote{Spheroidal coordinates have been employed. We
denote  $\rg=G\mbh/c^2\approx1.5\times10^{5}(\mbh/M_{\sun})\,$cm and we
use geometrized units with $c=G\mbh=1$ hereafter, which means that we scale
lengths with $\mbh$. Therefore, all quantities are dimensionless.} Clock-wise
distortion of the contours is due to frame-dragging near a rapidly
rotating Kerr black hole, and it is clearly visible in the Boyer-Lindquist
coordinates here. Notice that this dramatic distortion appears in the
graphical representation only. In order to achieve high accuracy of the 
tables the dragging effect has been largely eliminated in computations 
by means of appropriate transformation of coordinates, as described below.

\section{Photon flux from an accretion disc}
\begin{figure}[tbh]
 \begin{center}
  \begin{tabular}{ccc}
   a) \hspace{3.9cm} & b) \hspace{4.2cm} & c) \hspace{3cm} \\[-3mm]
   \includegraphics[height=2.5cm]{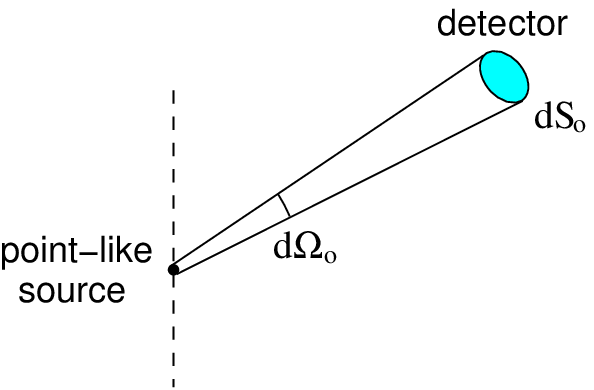} \hspace{2mm}
   & \includegraphics[height=2.5cm]{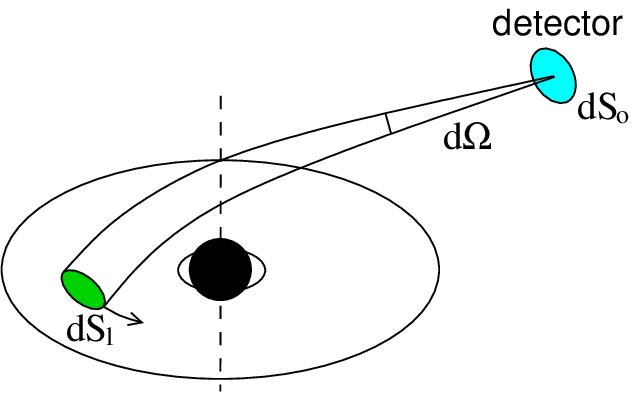} \hspace{2mm}
   & \includegraphics[height=2.5cm]{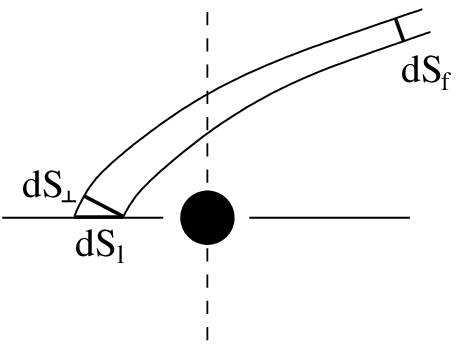}
  \end{tabular}
 \end{center}
\caption{Denomination of various elements of solid angles and areas
defined in the text: a) the light source appears to the observer to
be point-like;
b) the light rays received by the detector are  coming from different parts
of the disc (closer view of the disc than in previous figure);
c) area of a light tube changes as the light rays travel close to the black
hole (the disc is edge on).}
\label{denomination}
\end{figure}

Properties of radiation are described in terms of photon numbers. The
source appears as a point-like object for a distant observer, so that
the observer measures the flux entering the solid angle ${\rm d}
\Omega_{\rm o}$, which is associated with the detector area
${\rm d}S_{\rm o}{\equiv}D^2\,{\rm d}\Omega_{\rm o}$
(see Figure~\ref{denomination}a). This relation
defines distance $D$ between the observer and the source.
We denote the total photon flux received by a detector,
\begin{equation}
\label{flux}
N^{S}_{\rm{}o}(E)\equiv\frac{{\rm d}n(E)}{{\rm d}t\,{\rm d}S_{\rm o}}
={\int}{\rm d}\Omega\,N_{\loc}(E/g)\,g^2\,,
\end{equation}
where
\begin{equation}
N_{\loc}(E_{\loc})\equiv
\frac{{\rm d}n_{\loc}(E_{\loc})}{{\rm d}\tau\,{\rm d}S_{\loc}\,
{\rm d}\Omega_{\loc}}
\end{equation}
is a local photon flux emitted from the surface of the disc, ${\rm
d}n(E)$ is the number of photons with energy in the interval
${\langle}E,E+{\rm d}E\,\rangle$ and $g=E/E_{\loc}$ is the redshift
factor. The local flux, $N_\loc(E_\loc)$, may vary over the disc as well
as in time, and it can also depend on the local emission angle. This
dependency is emphasized explicitly only in the final formula
(\ref{emission}), otherwise it is omitted for brevity.

The emission arriving at the detector within the solid angle ${{\rm d}\Omega}$
(see Figure~\ref{denomination}b) originates from the proper area ${\rm
d}S_{\loc}$ on the disc (as measured in the rest frame co-moving with
the disc). Hence, in our computations we want to integrate the flux
contributions over a fine mesh on the disc surface. To achieve this aim, we
adjust eq.~(\ref{flux}) to the form
\begin{equation}
\label{N_o^S}
N^S_{\rm o}(E)=\frac{1}{D^2}\int {\rm d}S\,\frac{D^2{\rm d}\Omega}{{\rm d}S}\,
N_{\loc}(E/g)\,g^2=\frac{1}{D^2}\int {\rm d}S\,
\frac{{\rm d}S_{\perp}}{{\rm d}S}\,\frac{{\rm d}S_{\rm f}}{{\rm d}S_{\perp}}\,
N_{\loc}(E/g)\,g^2\, .
\end{equation}
Here ${{\rm{}d}S_{\rm f}}$ stands for an element of area perpendicular to light
rays corresponding to the solid angle ${\rm d}\Omega$ at a distance $D$,
${{\rm{}d}S_{\perp}}$ is the proper area measured in the local frame
of the disc and perpendicular to the
rays, and ${\rm d}S$ is the coordinate area for integration.
We integrate in a two-dimensional slice of a four-dimensional space-time, which
is specified by coordinates $\theta=\pi/2$ and $t=t_{\rm o}-\Delta t$ with
$\Delta t$ being a time delay with which photons from different parts of the
disc (that lies in the equatorial plane) arrive to the observer (at the same
coordinate time $t_{\rm o}$). Therefore,
let us define the coordinate area by (we employ coordinates
$t',\,\theta,\,r,\,\varphi$
with $t'=t-\Delta t$ and $\Delta t=\Delta t(r,\theta,\varphi)$)
\begin{equation}
\label{dS}
{\rm d}S\equiv|{\rm d}^2{\!S_{t'}}^{\theta}|=\left|\frac{\partial x^\mu}
{\partial t'}{\rm d}^2{\!S_{\mu}}^{\theta}\right|=|{\rm d}^2{\!S_t}^{\theta}|=
|g^{\theta\mu}{\rm d}^2\!S_{t\mu}|\, .
\end{equation}
We define the tensor ${\rm d}^2\!S_{\alpha\beta}$ by two four-vector elements
${\rm d}x_1^\mu\equiv({\rm d}t_1,{\rm d}r,0,0)$ and
${\rm d}x_2^\mu\equiv({\rm d}t_2,0,0,{\rm d}\varphi)$ and by
Levi-Civita tensor
$\varepsilon_{\alpha\beta\gamma\delta}$. The time components of these vectors,
${\rm d}t_1$ and ${\rm d}t_2$, are such that the vectors ${\rm d}x_1^\mu$ and
${\rm d}x_2^\mu$ lie in the tangent
space to the above defined space-time slice. Then we obtain
\begin{equation}
\label{dS2}
{\rm d}S=|g^{\theta\theta}\varepsilon_{t\theta\alpha\beta}\,{\rm d}x_1^{[\alpha}
{\rm d}x_2^{\beta]}|=g^{\theta\theta}\sqrt{-\|g_{\mu\nu}\|}\,{\rm d}r\,{\rm
d}\varphi={\rm d}r\,{\rm d}\varphi\, ,
\end{equation}
where $g_{\mu\nu}$ is the metric tensor and $\|g_{\mu\nu}\|$ is the
determinant of the metric. The proper area, ${\rm d}S_\perp$, perpendicular to
the light ray can be expressed covariantly in the following
way:
\begin{equation}
\label{dSperp}
{\rm d}S_\perp=-\frac{U^{\alpha}\,p^{\beta}\,{\rm d}^2\!S_{\alpha\beta}}
{U^{\mu}\,p_\mu}\, .
\end{equation}
Here, ${\rm d}S_\perp$ is the projection of an element of area,
defined by ${\rm d}^2\!S_{\alpha\beta}$, on a spatial slice of an observer
with velocity $U^\alpha$ and perpendicular to light rays.
$U^\alpha$ is four-velocity of an observer measuring the
area ${\rm d}S_\perp$, and $p^\beta$ is four-momentum of the photon.
The proper area ${\rm d}S_\perp$ corresponding to the same flux tube
is identical for all observers. This means that the last equation holds
true for any four-velocity $U^\alpha$, and we can express it as
\begin{equation}
\label{dS3}
p^{\beta}\,{\rm d}^2\!S_{\alpha\beta}+p_\alpha\,{\rm d}S_\perp = 0,\quad
\alpha=t,r,\theta,\varphi.
\end{equation}
For $\alpha=t$ (note that
${\rm d}^2\!S_{tr}={\rm d}^2\!S_{t\varphi}=0$) we get
\begin{equation}
\label{ratio_dS}
\frac{{\rm d}S_\perp}{{\rm d}S}=\left|\frac{1}{g^{\theta\theta}}
\frac{{\rm d}S_\perp}{{\rm d}^2\!S_{t\theta}}\right| =
\left|-\frac{p_\theta}{p_t}\right|=\frac{r\mu_{\rm e}}{g}\, .
\end{equation}
In the last equation we used the formula for the cosine of local emission
angle $\mu_{\rm e}$, see eq.~(\ref{cosine}), and the fact that we have chosen
such an affine parameter of the light geodesic that $p_t=-1$.
From eqs.~(\ref{N_o^S}), (\ref{dS}) and (\ref{ratio_dS}) we get for
the observed flux per unit solid angle
\begin{equation}
\label{emission1}
N^{\Omega}_{\rm o}(E)\equiv\frac{{\rm d}n(E)}{{\rm d}t\,{\rm d}\Omega_{\rm o}}=
{N_0}\int_{r_{\rm in}}^{r_{\rm out}}{\rm d}r\,\int_{\phi}^{\phi+{\Delta\phi}}
{\rm d}\varphi\,N_{\loc}(E/g)\,g\,l\,\mu_{\rm e}\,r,
\end{equation}
where $N_0$ is a normalization constant and
\begin{equation}
l=\frac{{\rm d}S_{\rm f}}{{\rm d}S_\perp}
\end{equation}
is the lensing factor in the limit $D\rightarrow\infty$
(the limit is performed while
keeping $D^2{\rm d}\Omega$ constant, see Figure~\ref{denomination}c).

For the line emission, the normalization constant $N_0$ is chosen in
such a way that the total flux from the disc is unity. In the case of a
continuum model, the flux is normalized to unity  at a certain value of the
observed energy (typically at $E=1$~keV, as in other {\sc{}xspec}
models).

Finally, the integrated flux per energy bin, $\Delta E$, is
\begin{eqnarray}
\label{emission}
\nonumber
\hspace*{-3mm}{\Delta}N^{\Omega}_{\rm o}(E,\Delta E,t) & 
\hspace*{-3mm} = &\hspace*{-3mm}
\int_{E}^{E+\Delta E}{\rm d}\bar{E}\,N^{\Omega}_{\rm o}(\bar{E},t)=\\
& \hspace*{-3mm} = & \hspace*{-3mm}N_0\int_{r_{\rm in}}^{r_{\rm out}}\!\!{\rm d}r\,
\int_{\phi}^{\phi+{\Delta\phi}}\!\!{\rm d}\varphi\,\int_{E/g}^{(E+\Delta E)/g}
\!\!{\rm d}E_{\loc}\,N_{\loc}(E_{\loc},r,\varphi,\mu_{\rm e},t-\Delta t)\,g^2\,l\,
\mu_{\rm e}\,r\, ,\hspace{2mm}
\end{eqnarray}
where $\Delta t$ is the relative time delay with which photons arrive to the
observer from different parts of the disc. The transfer functions
$g,\,l,\,\mu_{\rm e}$ and $\Delta t$ are read from the FITS file {\tt
KBHtablesNN.fits} described in Appendix~\ref{appendix3a}. This equation
is numerically integrated for a given local flux
$N_\loc(E_{\loc},r,\varphi,\mu_{\rm e},t-\Delta t)$ in all hereby
described new general relativistic {\it non-axisymmetric models}.

Let us assume that the local emission is stationary and the dependence on the
axial coordinate is only through the prescribed dependence on the local emission
angle $f(\mu_{\rm e})$ (limb darkening/brightening law) together with an
arbitrary radial dependence $R(r)$, i.e.
\begin{equation}
N_\loc(E_{\loc},r,\varphi,\mu_{\rm e},t-\Delta t)\equiv N_\loc(E_\loc)\,R(r)\,
f(\mu_{\rm e}).
\end{equation}
The observed flux $N_{\rm o}^{\Omega}(E)$
is in this case given by 
\begin{equation}
N_{\rm o}^{\Omega}(E)=\int_{-\infty}^{\infty}{\rm
d}E_\loc\,N_\loc(E_\loc)\,G(E,E_\loc),
\end{equation}
where
\begin{equation}
G(E,E_\loc)=N_0\int_{r_{\rm in}}^{r_{\rm out}}{\rm
d}r\,R(r)\int_{0}^{2\pi}{\rm d}\varphi\,f(\mu_{\rm e})\,g^2\,l\,\mu_{\rm
e}\,r\,\delta(E-gE_\loc).
\end{equation}
In this case, the integrated flux can be expressed in the following way:
\begin{eqnarray}
\nonumber
{\Delta}N^{\Omega}_{\rm o}(E,\Delta E) & = & \int_{E}^{E+\Delta E}{\rm d}\bar{E}\,
N^{\Omega}_{\rm o}(\bar{E})\quad=\hspace{7cm}\\
\nonumber
& & \hspace*{-2.8cm} = \int_{E}^{E+\Delta E}{\rm d}\bar{E}\,
{N_0}\int_{r_{\rm in}}^{r_{\rm out}}{\rm d}r\,R(r)\,\int_{0}^{2\pi}
{\rm d}\varphi\,f(\mu_{\rm e})\,N_{\loc}(\bar{E}/g)\,g\,l\,\mu_{\rm e}\,r\,
\int_{-\infty}^{\infty}{\rm d}E_\loc\,\delta(E_\loc-\bar{E}/g)=\\
\label{axisym_emission}
& & \hspace*{-2.8cm} = N_0\int_{r_{\rm in}}^{r_{\rm out}}{\rm d}r\,R(r)\,
\int_{-\infty}^{\infty}{\rm d}E_{\loc}\,N_{\loc}(E_{\loc})
\int_{E/E_{\loc}}^{(E+\Delta E)/E_{\loc}}{\rm d}\bar{g}\,F(\bar{g})\, ,
\end{eqnarray}
where we substituted $\bar{g}=\bar{E}/E_\loc$ and
\begin{equation}
\label{conv_function}
F(\bar{g}) = \int_{0}^{2\pi}{\rm d}\varphi\,f(\mu_{\rm e})\,g^2\,l\,
\mu_{\rm e}\,r\,\delta(\bar{g}-g)\, .
\end{equation}
Eq.~(\ref{axisym_emission}) is numerically integrated in all
{\it axially symmetric models}. The function
${\rm d}F(\bar{g})\equiv {\rm d}\bar{g}\,F(\bar{g})$ has been
pre-calculated for several limb darkening/brightening laws
$f(\mu_{\rm e})$ and stored in separate files,
{\tt KBHlineNN.fits} (see Appendix~\ref{appendix3b}).

\section{Stokes parameters in strong gravity regime}

\begin{figure}[tbh]
 \begin{center}
  \begin{tabular}{c@{\quad}l}
   \includegraphics[height=5cm]{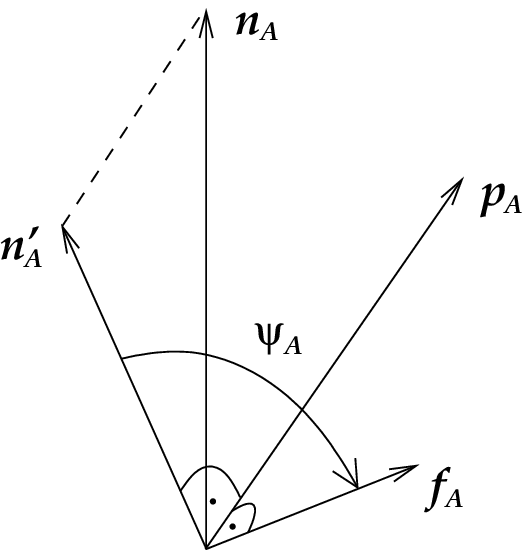} &
   \parbox[t]{9cm}{\vspace*{-5.5cm}
     \begin{itemize} \itemsep -2pt
       \item [\rm(i)] Let three-vectors $\vec{p_A}$, $\vec{n_A}$, $\vec{n'_A}$
and $\vec{f_A}$ be the momentum of a photon, normal to
the disc, projection of the normal to the plane perpendicular to the momentum
and a vector which is parallelly transported along the geodesic (as
four-vector), respectively;
       \item [\rm (ii)] let $\Psi_A$ be an angle between $\vec{n'_A}$
and $\vec{f_A}$;
       \item [\rm (iii)] let the quantities in (i) and (ii) be evaluated at
the disc for $A=1$ with respect to the local rest frame co-moving with the
disc, and at infinity for \hbox{$A=2$} with respect to the stationary observer
at the same light geodesic;
       \item [\rm (iv)] then the change of polarization angle is defined
as $\Psi=\Psi_2-\Psi_1$.
     \end{itemize}}
  \end{tabular}
 \end{center}
\caption{Definition of the change of polarization angle $\Psi$.}
\label{pol_angle}
\end{figure}

For polarization studies, Stokes parameters are used. Let us define specific
Stokes parameters in the following way:
\begin{equation}
i_\nu\equiv \frac{I_{\nu}}{E}\, ,\quad q_\nu\equiv \frac{Q_{\nu}}{E}\, ,\quad
u_\nu\equiv \frac{U_{\nu}}{E}\, ,\quad v_\nu\equiv \frac{V_{\nu}}{E}\, ,
\end{equation}
where $I_{\nu}$, $Q_{\nu}$, $U_{\nu}$ and $V_{\nu}$ are Stokes
parameters for light with frequency $\nu$, $E$ is energy of a photon at
this frequency. Further on, we drop the index $\nu$ but we will always
consider these quantities for light of a given frequency. We can calculate
the integrated specific Stokes parameters (per energy bin), i.e.\ $\Delta
i_{\rm o}$, $\Delta q_{\rm o}$, $\Delta u_{\rm o}$ and $\Delta v_{\rm
o}$. These are the quantities that the observer determines from the local
specific Stokes parameters $i_\loc$, $q_\loc$, $u_\loc$ and $v_\loc$ on the disc
in the following way:
\begin{eqnarray}
\label{S1}
{\Delta}i_{\rm o}(E,\Delta E) & = & N_0\int{\rm d}S\,\int{\rm d}E_{\loc}\,
i_{\loc}(E_{\loc})\,Fr\, ,\\
\label{S2}
{\Delta}q_{\rm o}(E,\Delta E) & = & N_0\int{\rm d}S\,\int{\rm d}E_{\loc}\,
[q_{\loc}(E_{\loc})\cos{2\Psi}-u_{\loc}(E_{\loc})\sin{2\Psi}]\,Fr\, ,\\
\label{S3}
{\Delta}u_{\rm o}(E,\Delta E) & = & N_0\int{\rm d}S\,\int{\rm d}E_{\loc}\,
[q_{\loc}(E_{\loc})\sin{2\Psi}+u_{\loc}(E_{\loc})\cos{2\Psi}]\,Fr\, ,\\
\label{S4}
{\Delta}v_{\rm o}(E,\Delta E) & = & N_0\int{\rm d}S\,\int{\rm d}E_{\loc}\,
v_{\loc}(E_{\loc})\,Fr\, .
\end{eqnarray}
Here, $F\equiv F(r,\varphi)=g^2\,l\,\mu_{\rm e}$ is a transfer
function, $\Psi$ is the angle by which a vector parallelly transported
along the light geodesic rotates. We refer to this angle also as a
change of the polarization angle, because the polarization vector is parallelly
transported along light geodesics. See Figure~\ref{pol_angle} for an exact
definition of the angle $\Psi$. The integration boundaries are the
same as in eq.~(\ref{emission}). As can be seen from the
definition, the first specific Stokes parameter is equal to the photon
flux, therefore, eqs.~(\ref{emission}) and (\ref{S1}) are
identical. The local specific Stokes parameters may depend on $r$,
$\varphi$, $\mu_{\rm e}$ and $t-\Delta t$, which we did not state in the
eqs.~(\ref{S1})--(\ref{S4}) explicitly for simplicity.

The specific Stokes parameters that the observer measures may vary in time
in the case when the local parameters also depend on time.
In eqs.~(\ref{S1})--(\ref{S4}) we used a law of transformation of the Stokes
parameters by the rotation of axes (eqs.~(I.185) and (I.186) in
\cite{chandrasekhar_1960}).

An alternative way for expressing polarization of light is by using the
degree of polarization $P_{\rm o}$ and two polarization angles
$\chi_{\rm o}$ and $\xi_{\rm o}$, defined by
\begin{eqnarray}
P_{\rm o} & = & \sqrt{q_{\rm o}^2+u_{\rm o}^2+v_{\rm o}^2}/i_{\rm o}\, , \\
\tan{2\chi_{\rm o}} & = & u_{\rm o}/q_{\rm o}\, ,\\
\sin{2\xi_{\rm o}} & = & v_{\rm o}/\sqrt{q_{\rm o}^2+u_{\rm o}^2+v_{\rm o}^2}\, .
\end{eqnarray}

\section{New model for {\fontfamily{phv}\fontshape{sc}\selectfont xspec}}
We have developed several general relativistic models for line emission
and Compton reflection continuum. The line models are supposed to be
more accurate and versatile than the {\sc laor} model \cite{laor_1991}, and
substantially faster than the {\sc kerrspec} model \cite{martocchia00}.
Several models of intrinsic emissivity were employed, including the lamp-post
model \cite{matt92}.  Among other features,
these models allow various parameters to be fitted such as the black-hole
angular momentum, observer inclination, accretion disc size and some of the
parameters characterizing disc emissivity and primary illumination
properties. They also
allow a change in the grid resolution and, hence, to control accuracy and
computational speed. Furthermore, we developed very general
convolution models. All these models are based on pre-calculated tables
described in Section~\ref{transfer_functions} and thus the geodesics
do not need to be calculated each time one integrates the disc emission.
These tables are calculated for the vacuum Kerr space-time and for a Keplerian
co-rotating disc plus matter that is freely falling below the marginally
stable orbit. The falling matter has the energy and angular momentum of
the matter at the marginally stable orbit. It is possible to use
different pre-calculated tables if they are stored in a specific FITS
file (see Appendix~\ref{appendix3a} for its detailed description).

There are two types of new models. The first type integrates
the local disc emission in both of the polar coordinates on the disc and thus
enables one to choose non-axisymmetric area of integration. This 
option is useful for example when computing spectra
of spots or partially obscured discs. One can also choose the
resolution of integration and thus control the precision and speed of
the computation. The second type of models is axisymmetric -- the
axially dependent part of the emission from rings is pre-calculated and
stored in a FITS file (the function ${\rm d}F(\bar{g})={\rm
d}\bar{g}\,F(\bar{g})$ from (\ref{conv_function}) is integrated for
different radii with the angular grid having $20\,000$ points). These
models have less
parameters that can be fitted and thus are less flexible even though more
suited to the standard analysis approach. On the other
hand they are fast because the emission is integrated only in one
dimension (in the radial coordinate of the disc). It may be worth emphasizing
that the assumption about axial symmetry concerns only the form of intrinsic
emissivity of the disc, which cannot depend on the polar angle in this case, not
the shape of individual light rays, which is always
complicated near a rotating black hole.

\begin{table}[tbh]
\begin{center}
\begin{tabular}[t]{l|c|c|c|c}
parameter & unit & default value & minimum value & maximum value \\ \hline
\hspace*{0.5em}{\tt a/M}       & $GM/c$   & 0.9982 &  0.      & 1.    \\
\hspace*{0.5em}{\tt theta\_o}  & deg      & 30.    &  0.      & 89.   \\
\hspace*{0.5em}{\tt rin-rh}    & $GM/c^2$ & 0.     &  0.      & 999.  \\
\hspace*{0.5em}{\tt ms}        & --       & 1.     &  0.      & 1.    \\
\hspace*{0.5em}{\tt rout-rh}   & $GM/c^2$ & 400.   &  0.      & 999.  \\
\hspace*{0.5em}{\tt zshift}    & --       & 0.     &  -0.999. & 10.   \\
\hspace*{0.5em}{\tt ntable}    & --       & 0.     &  0.      & 99.   \\
\end{tabular}
\end{center}
\caption{Common parameters for all models.}
\label{common_par1}
\end{table}

There are several parameters and switches that are common for all new models
(see Table~\ref{common_par1}):
\begin{description} \itemsep -2pt
 \item[{\tt a/M}] -- specific angular momentum of the Kerr black hole in units
 of $GM/c$ ($M$ is the mass of the central black hole),
 \item [{\tt theta\_o}] -- the inclination of the observer in degrees,
 \item [{\tt rin-rh}] -- inner radius of the disc relative to the black-hole
 horizon in units of $GM/c^2$,
 \item [{\tt ms}] -- switch for the marginally stable orbit,
 \item [{\tt rout-rh}] -- outer radius of the disc relative to the black-hole
 horizon in units of $GM/c^2$,
 \item [{\tt zshift}] -- overall redshift of the object,
 \item [{\tt ntable}] -- number of the FITS file with pre-calculated tables to
 be used.
\end{description}
  The inner and outer radii are given relative to the black-hole horizon and,
therefore, their minimum value is zero. This becomes handy when one fits the
{\tt a/M} parameter, because the horizon of the black hole as well as the
marginally stable orbit changes with {\tt a/M}, and so the lower limit for
inner and outer disc edges
cannot be set to constant values. The {\tt ms} switch determines whether
we intend to integrate also emission below the marginally stable orbit.
If its value is set to zero and the inner radius of the disc is below this
orbit then the emission below the marginally stable orbit is taken
into account, otherwise it is not.

The {\tt ntable} switch determines which of the pre-calculated tables
should be used for intrinsic emissivity.
In particular, ${\tt ntable}=0$ for {\tt KBHtables00.fits}
({\tt KBHline00.fits}), ${\tt ntable}=1$ for {\tt KBHtables01.fits}
({\tt KBHline01.fits}), etc., corresponding to non-axisymmetric
(axisymmetric) models.

\begin{figure}[tbh]
\begin{center}
\includegraphics[width=5.3cm]{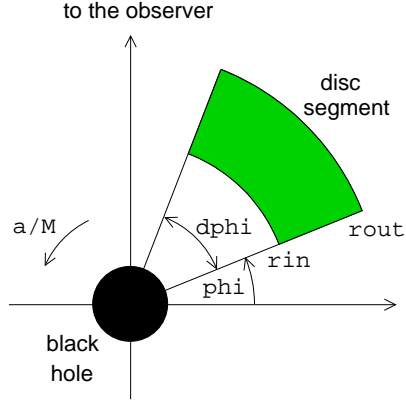}
\end{center}
\caption{Segment of a disc from which emission comes (view from above).}
\label{sector}
\end{figure}

\begin{table}[tbh]
\begin{center}
\begin{tabular}[h]{l|c|c|c|c}
parameter & unit & default value  & minimum value & maximum value \\ \hline
\hspace*{0.5em}{\tt phi}      & deg & 0.   &  -180. & 180.       \\
\hspace*{0.5em}{\tt dphi}     & deg & 360. &  0.    & 360.       \\
\hspace*{0.5em}{\tt nrad}     & --  & 200. &  1.    & 10000.     \\
\hspace*{0.5em}{\tt division} & --  & 1.   &  0.    & 1.         \\
\hspace*{0.5em}{\tt nphi}     & --  & 180. &  1.    & 20000.     \\
\hspace*{0.5em}{\tt smooth}   & --  & 1.   &  0.    & 1.         \\
\hspace*{0.5em}{\tt Stokes}   & --  & 0.   &  0.    & 6.         \\
\end{tabular}
\end{center}
\caption{Common parameters for non-axisymmetric models.}
\label{common_par2}
\end{table}
The following set of parameters is relevant only for non-axisymmetric models
(see Table~\ref{common_par2}):
\begin{description} \itemsep -2pt
 \item[{\tt phi}] -- position angle of the axial sector of the disc in degrees,
 \item[{\tt dphi}] -- inner angle of the axial sector of the disc in degrees,
 \item[{\tt nrad}] -- radial resolution of the grid,
 \item[{\tt division}] -- switch for spacing of radial grid
 ($0$ -- equidistant, $1$ -- exponential),
 \item[{\tt nphi}] -- axial resolution of the grid,
 \item[{\tt smooth}] -- switch for performing simple smoothing
 ($0$ -- no, $1$ -- yes),
 \item[{\tt Stokes}] -- switch for computing polarization
 (see Table~\ref{stokes}).
\end{description}
  The {\tt phi} and {\tt dphi} parameters determine the axial sector of the disc
from which emission comes (see Figure~\ref{sector}). The {\tt nrad} and {\tt nphi}
parameters determine the grid for numerical integration.
If the {\tt division} switch is zero, the radial grid is equidistant;
if it is equal to unity then the radial grid is exponential
(i.e.\ more points closer to the black hole).

If the {\tt smooth} switch is set to unity then a simple smoothing
is applied to the final spectrum. Here $N_{\rm o}^{\Omega}(E_{\loc})=
[N_{\rm o}^{\Omega}(E_{\rm i-1})+2N_{\rm o}^{\Omega}(E_{\loc})+
N_{\rm o}^{\Omega}(E_{\rm i+1})]/4$.

If the {\tt Stokes} switch is different from zero, then the model also
calculates polarization. Its value determines, which of the Stokes parameters
should be computed by {\sc{}xspec}, i.e.\ what will be stored in the output
array for the photon flux {\tt photar}; see Table~\ref{stokes}.
(If ${\tt Stokes}\neq0$ then a new {\sf ascii} data file {\tt stokes.dat} is
created in the current directory,
where values of energy $E$ together with all Stokes parameters
$i,\,q,\,u,\,v,\,P,\,\chi$[deg] and $\xi$[deg] are stored, each in one
column.)

A realistic model of polarization has been currently implemented only in the
{\sc kyl1cr} model (see Section~\ref{kyl1cr}
below). In other models, a simple assumption is made -- the local emission
is assumed to be linearly polarized in the direction perpendicular to
the disc (i.e.\ $q_{\loc}=i_{\loc}=N_\loc$ and $u_\loc=v_\loc=0$). In all
models (including {\sc kyl1cr}) there is always no final circular polarization
(i.e.\ $v=\xi=0$), which follows from the fact that the fourth local
Stokes parameter is zero in each model.

\begin{table}[tbh]
\begin{center}
\begin{tabular*}{10.5cm}{c|l}
value & photon flux array {\tt photar} contains$^{\dag}{}^{\ddag}$\\ \hline
0 & $i=I/E$, where $I$ is the first Stokes parameter (intensity) \\
1 & $q=Q/E$, where $Q$ is the second Stokes parameter \\
2 & $u=U/E$, where $U$ is the third Stokes parameter \\
3 & $v=V/E$, where $V$ is the fourth Stokes parameter \\
4 & degree of polarization, $P=\sqrt{q^2+u^2+v^2}/i$ \\
5 & angle $\chi$[deg] of polarization, $\tan{2\chi}=u/q$\\
6 & angle $\xi$[deg] of polarization, $\sin{2\xi}=v/\sqrt{q^2+u^2+v^2}$\\ \hline
\multicolumn{2}{l}{\parbox{10.cm}{\vspace*{1mm}\footnotesize
$^{\dag}$~the {\tt photar} array contains values described in the table
and multiplied by width of the corresponding energy bin\\
$^{\ddag}~E$ is energy of observed photons}}\\
\end{tabular*}
\end{center}
\caption{Definition of the {\tt Stokes} parameter.}
\label{stokes}
\end{table}

\section{Models for a relativistic spectral line}
Three general relativistic line models are included in the new set of
{\sc{}xspec} routines -- non-axisymmetric Gaussian line model
{\sc kyg1line}, axisymmetric Gaussian line model {\sc kygline} and
fluorescent lamp-post line model {\sc kyf1ll}.

\subsection{Non-axisymmetric Gaussian line model
{\fontfamily{phv}\fontshape{sc}\selectfont kyg1line}}
The {\sc kyg1line} model computes the integrated flux from the
disc according to eq.~(\ref{emission}). It assumes that the local
emission from the disc is
\begin{eqnarray}
\label{line_emiss1a}
N_{\loc}(E_\loc) & = & \frac{1}{r^{\tt alpha}}\,f(\mu_{\rm e})\,\exp{\left
[-\left (1000\,\frac{E_\loc-{\tt Erest}}{\sqrt{2}\,{\tt sigma}}\right )^2
\right ]} \quad {\rm for}\quad r\ge r_{\rm b}\, ,\\
\label{line_emiss1b}
N_{\loc}(E_\loc) & = & {\tt jump}\ r_{\rm b}^{{\tt beta} - {\tt alpha}}\,
\frac{1}{r^{\tt beta}}\,f(\mu_{\rm e})\,\exp{\left [-\left (1000\,
\frac{E_\loc-{\tt Erest}}{\sqrt{2}\,{\tt sigma}}\right )^2\right ]} \quad
{\rm for}\quad r<r_{\rm b}\, .\hspace{5mm}
\end{eqnarray}
The local emission is assumed to be a Gaussian line with its peak flux
depending on the radius
as a broken power law. The line is defined by nine points equally spaced with
the central point at its maximum.
The normalization constant $N_0$ in (\ref{emission}) is such that the total
integrated flux of the line is unity.
The parameters defining the Gaussian line are (see Table~\ref{kyg1line_par}):
\begin{description} \itemsep -2pt
 \item[{\tt Erest}] -- rest energy of the line in keV,
 \item[{\tt sigma}] -- width of the line in eV,
 \item[{\tt alpha}] -- radial power-law index for the outer region,
 \item[{\tt beta}] -- radial power-law index for the inner region,
 \item[{\tt rb}] -- parameter defining the border between regions with different
 power-law indices,
 \item[{\tt jump}] -- ratio between flux in the inner and outer regions at
 the border radius,
 \item[{\tt limb}] -- switch for different limb darkening/brightening laws.
\end{description}
There are two regions with different power-law dependences with indices
{\tt alpha} and {\tt beta}. The power law changes at the border radius
$r_{\rm b}$ where the local emissivity does not need to be continuous
(for ${\tt jump}\neq 1$). The {\tt rb} parameter defines this radius in the
following way:
\begin{eqnarray}
\label{rb1}
r_{\rm b} &=& {\tt rb} \times r_{\rm ms} \quad {\rm for}\quad {\tt rb}\ge
0\, ,\\
\label{rb2}
r_{\rm b} &=& -{\tt rb}+r_{\rm h} \quad {\rm for}\quad {\tt rb}< 0\, ,
\end{eqnarray}
where $r_{\rm ms}$ is the radius of the marginally stable orbit and $r_{\rm h}$
is the radius of the horizon of the black hole.

\begin{table}[tbh]
\begin{center}
\begin{tabular}[h]{r@{}l|l|c|c|c}
\multicolumn{2}{c|}{parameter} & unit & default value  & minimum value &
maximum value \\ \hline
&{\tt a/M}       & $GM/c$   & 0.9982  & 0.       & 1.        \\
&{\tt theta\_o}  & deg      & 30.     & 0.       & 89.       \\
&{\tt rin-rh}    & $GM/c^2$ & 0.      & 0.       & 999.      \\
&{\tt ms}        & --       & 1.      & 0.       & 1.        \\
&{\tt rout-rh}   & $GM/c^2$ & 400.    & 0.       & 999.      \\
&{\tt phi}       & deg      & 0.      & -180.    & 180.      \\
&{\tt dphi}      & deg      & 360.    & 0.       & 360.      \\
&{\tt nrad}      & --       & 200.    & 1.       & 10000.    \\
&{\tt division}  & --       & 1.      & 0.       & 1.        \\
&{\tt nphi}      & --       & 180.    & 1.       & 20000.    \\
&{\tt smooth}    & --       & 1.      & 0.       & 1.        \\
&{\tt zshift}    & --       & 0.      & -0.999   & 10.       \\
&{\tt ntable}    & --       & 0.      & 0.       & 99.       \\
{*}&{\tt Erest}  & keV      & 6.4     & 1.       & 99.       \\
{*}&{\tt sigma}  & eV       & 2.      & 0.01     & 1000.     \\
{*}&{\tt alpha}  & --       & 3.      & -20.     & 20.       \\
{*}&{\tt beta}   & --       & 4.      & -20.     & 20.       \\
{*}&{\tt rb}     & $r_{\rm ms}$ & 0.  & 0.       & 160.      \\
{*}&{\tt jump}   & --       & 1.      & 0.       & 1e6       \\
{*}&{\tt limb}   & --       & -1.     & -10.     & 10.       \\
&{\tt Stokes}    & --       & 0.      & 0.       & 6.        \\
\end{tabular}
\end{center}
\caption{Parameters of the non-axisymmetric Gaussian line model {\sc{}kyg1line}.
Model parameters that are not common for all non-axisymmetric models are
denoted by asterisk.}
\label{kyg1line_par}
\end{table}

The function $f(\mu_{\rm e})=f(\cos{\,\delta_{\rm e}})$ in (\ref{line_emiss1a})
and (\ref{line_emiss1b}) describes the limb darkening/brightening law, i.e.\
the dependence of the local emission on the local emission angle. Several limb
darkening/brightening laws are implemented:
\begin{eqnarray}
\label{isotropic}
f(\mu_{\rm e}) & = & 1 \quad  {\rm for}  \quad {\tt limb}=0\, , \\
\label{laor}
f(\mu_{\rm e}) & = & 1 + 2.06 \mu_{\rm e} \quad  {\rm for}  \quad {\tt limb}=
-1\, ,\\
\label{haardt}
f(\mu_{\rm e}) & = & \ln{(1+\mu_{\rm e}^{-1})} \quad  {\rm for}  \quad {\tt limb}=
-2\, ,\\
\label{other_limb}
f(\mu_{\rm e}) & = & \mu_{\rm e}^{\tt limb} \quad {\rm for} \quad {\tt limb}
\ne 0,-1,-2\, .
\end{eqnarray}
Eq.~(\ref{isotropic}) corresponds to the isotropic local emission,
eq.~(\ref{laor}) corresponds to
limb darkening in an optically thick electron scattering atmosphere
(used by Laor \cite{laor_1990,laor_1991,phillips_1986}),
and eq.~(\ref{haardt}) corresponds
to limb brightening predicted by some models of a fluorescent line emitted
by an accretion disc due to X-ray illumination
\cite{ghisellini_1994,haardt_1993}.

There is also a similar model {\sc kyg2line} present among the new {\sc{} xspec}
models, which is useful when fitting two general relativistic lines
simultaneously. The parameters are the same as in the {\sc kyg1line} model
except that there are two sets of those parameters
describing the local Gaussian line emission. There is one more parameter
present, {\tt ratio21}, which is the ratio of the maximum of the second local
line to the maximum of the first local line. Polarization computations are
not included in this model.

\subsection{Axisymmetric Gaussian line model
{\fontfamily{phv}\fontshape{sc}\selectfont kygline}}
This model uses eq.~(\ref{axisym_emission}) for computing the
disc emission with local flux being
\begin{eqnarray}
N_\loc(E_\loc) & = & \delta(E_\loc-{\tt Erest})\, ,\\
R(r) & = & r^{-{\tt alpha}}\, .
\end{eqnarray}
The function ${\rm d}F(\bar{g})\equiv {\rm d}\bar{g}\,F(\bar{g})$ in
(\ref{conv_function}) was pre-calculated for three different limb
darkening/brightening laws (\ref{isotropic}) -- (\ref{haardt}) and stored in
corresponding FITS files {\tt KBHline00.fits} -- {\tt KBHline02.fits}. The local
emission is a delta function with its maximum depending on the radius as a power
law with index {\tt alpha} and also depending on the local emission angle. The
normalization constant $N_0$ in (\ref{axisym_emission}) is such that the total
integrated flux of the line is unity.

\begin{table}[tbh]
\begin{center}
\begin{tabular}[h]{r@{}l|c|c|c|c}
\multicolumn{2}{c|}{parameter} & unit & default value  & minimum value &
maximum value \\ \hline
&{\tt a/M}      & $GM/c$   & 0.9982  & 0.     & 1.   \\
&{\tt theta\_o} & deg      & 30.     & 0.     & 89.  \\
&{\tt rin-rh}   & $GM/c^2$ & 0.      & 0.     & 999. \\
&{\tt ms}       & --       & 1.      & 0.     & 1.   \\
&{\tt rout-rh}  & $GM/c^2$ & 400.    & 0.     & 999. \\
&{\tt zshift}   & --       & 0.      & -0.999 & 10.  \\
&{\tt ntable}   & --       & 1.      & 0.     & 99.  \\
{*}&{\tt Erest} & keV      & 6.4     & 1.     & 99.  \\
{*}&{\tt alpha} & --       & 3.      & -20.   & 20.  \\
\end{tabular}
\end{center}
\caption{Parameters of the axisymmetric Gaussian line model {\sc{}kygline}.
Model parameters that are not common for all axisymmetric models are denoted
by asterisk.}
\label{kygline_par}
\end{table}
There are less parameters defining the line in this model than in
the previous one (see Table~\ref{kygline_par}):
\begin{description} \itemsep -2pt
 \item[{\tt Erest}] -- rest energy of the line in keV,
 \item[{\tt alpha}] -- radial power-law index.
\end{description}
Note that the limb darkening/brightening law can be chosen by means of
the {\tt ntable} switch.

This model is much faster than the non-axisymmetric {\sc kyg1line} model.
Although it is not possible to change the resolution grid on the disc,
it is hardly needed because the resolution is set to be very large,
corresponding to ${\tt nrad}=500$,
${\tt division}=1$ and ${\tt nphi}=20\,000$ in the {\sc kyg1line} model,
which is more than sufficient in most cases.
(These values apply if the maximum range of radii is selected,
i.e.\ {\tt rin}=0, {\tt ms}=0 and {\tt rout}=999; in case of
a smaller range the number of points decreases accordingly.)
This means that the resolution of the {\sc kyg1line} model is much
higher than what can be achieved with the {\tt laor} model, and the
performance is still very good.

\subsection{Non-axisymmetric fluorescent lamp-post line model
{\fontfamily{phv}\fontshape{sc}\selectfont kyf1ll}}
\label{section_kyf1ll}
\begin{figure}[tbh]
 \begin{center}
  \begin{tabular}{cc}
   a) \hspace{5.8cm} & b) \hspace{5.8cm} \\[-3mm]
   \includegraphics[height=3cm]{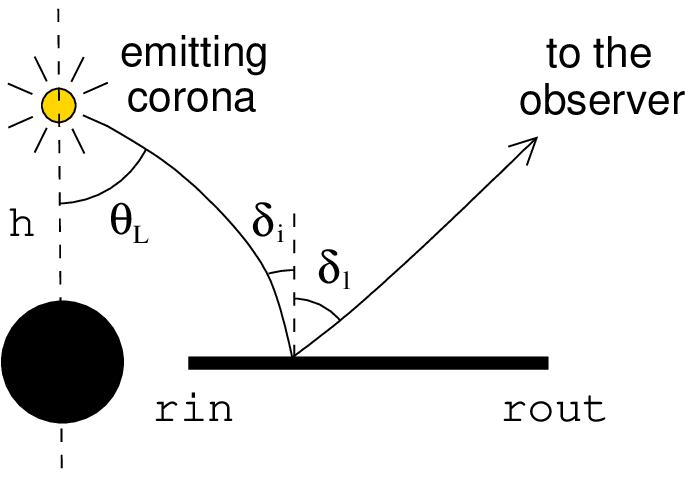} \hspace{2mm}
   & \includegraphics[height=3cm]{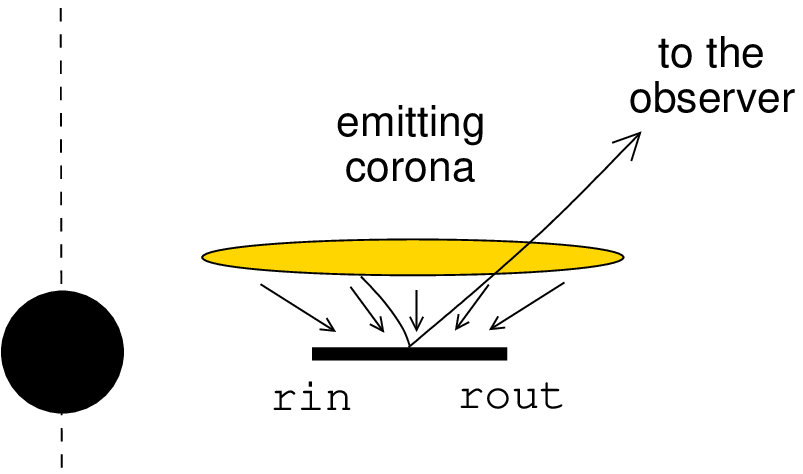}
  \end{tabular}
 \end{center}
\caption{Reflection models: a) lamp-post model; b) diffuse corona model.}
\label{compton_reflection}
\end{figure}

The line in this model is induced by the illumination of the disc from the
primary power-law source
located on the axis at {\tt height} above the black hole.
This model computes the final spectrum according to eq.~(\ref{emission})
with the local photon flux
\begin{eqnarray}
\nonumber
N_{\loc}(E_\loc) & = & g_{\rm L}^{{\tt PhoIndex}-1}\frac{\sin\theta_{\rm L}
{\rm d}\theta_{\rm L}}{r\,{\rm d}r}\,\sqrt{1-\frac{2{\tt height}}
{{\tt height}^2+a^2}}\,f(\mu_{\rm i},\mu_{\rm e})\\
\label{fl_emission}
 & & \times\ \exp{\left [-\left
(1000\,\frac{E_\loc-{\tt Erest}}{\sqrt{2}\,{\tt sigma}}\right )^2\right ]}.
\end{eqnarray}
Here, $g_{\rm L}$ is ratio of the energy of a photon received by the
accretion disc to the energy of the same photon when emitted from a source
on the axis, $\theta_{\rm L}$ is an angle under which the photon is emitted
from the source (measured in the local frame of the source) and
$\mu_{\rm i}\equiv\cos{\,\delta_{\rm i}}$ is the cosine of the incident angle
(measured in the local frame of the disc) -- see
Figure~\ref{compton_reflection}.
All of these functions depend on {\tt height} above the black hole at which
the source is located and on the rotational parameter {\tt a/M} of the black
hole. Values of $g_{\rm L}$, $\theta_{\rm L}$ and $\mu_{\rm i}$ for a given
height and rotation are read from the lamp-post tables {\tt lamp.fits}
(see Appendix~\ref{appendix3c}). At present, only tables for
${\tt a/M}=0.9987492$ (i.e.\ for the horizon of the black hole $r_{\rm h}=1.05$)
 and ${\tt height}=$ $2$,$\,3,\,4,\,5,\,6,\,8,\,10,\,12$,$\,15,\,20,\,30,\,50,\,75$ and
$100$ are included in {\tt lamp.fits},
therefore, the {\tt a/M} parameter is used only for the negative values
of {\tt height} (see below).

The factor in front of the function $f(\mu_{\rm i},\mu_{\rm e})$ gives
the radial dependence of the disc emissivity, which is different from the
assumed broken power law in the {\sc kyg1line} model.
For the derivation of this factor, which characterizes the illumination from
a primary source on the axis see Appendix~\ref{appendix2}.

\begin{table}[tbh]
\begin{center}
\begin{tabular}[h]{r@{}l|c|c|c|c}
\multicolumn{2}{c|}{parameter} & unit & default value  & minimum value &
maximum value \\ \hline
&{\tt a/M}         & $GM/c$   & 0.9982  & 0.       & 1.        \\
&{\tt theta\_o}    & deg      & 30.     & 0.       & 89.       \\
&{\tt rin-rh}      & $GM/c^2$ & 0.      & 0.       & 999.      \\
&{\tt ms}          & --       & 1.      & 0.       & 1.        \\
&{\tt rout-rh}     & $GM/c^2$ & 400.    & 0.       & 999.      \\
&{\tt phi}         & deg      & 0.      & -180.    & 180.      \\
&{\tt dphi}        & deg      & 360.    & 0.       & 360.      \\
&{\tt nrad}        & --       & 200.    & 1.       & 10000.    \\
&{\tt division}    & --       & 1.      & 0.       & 1.        \\
&{\tt nphi}        & --       & 180.    & 1.       & 20000.    \\
&{\tt smooth}      & --       & 1.      & 0.       & 1.        \\
&{\tt zshift}      & --       & 0.      & -0.999   & 10.       \\
&{\tt ntable}      & --       & 0.      & 0.       & 99.       \\
{*}&{\tt PhoIndex} & --       & 2.      & 1.5      & 3.        \\
{*}&{\tt height}   & $GM/c^2$ & 3.      & -20.     & 100.      \\
{*}&{\tt Erest}    & keV      & 6.4     & 1.       & 99.       \\
{*}&{\tt sigma}    & eV       & 2.      & 0.01     & 1000.     \\
&{\tt Stokes}      & --       & 0.      & 0.       & 6.        \\
\end{tabular}
\end{center}
\caption{Parameters of the fluorescent lamp-post line model {\sc{}kyf1ll}.
Model parameters that are not common for all non-axisymmetric models are
denoted by asterisk.}
\label{kyf1ll_par}
\end{table}

The function $f(\mu_{\rm i},\mu_{\rm e})$ is a coefficient of
reflection. It depends on the incident and reflection angles.
Although the normalization of this function also depends on the photon
index of the power-law emission from a primary source,
we do not need to take this into account because the final spectrum
is always normalized to unity. Values of this function are read
from a pre-calculated table which is stored in
{\tt fluorescent\_line.fits} file (see \cite{matt_1991} and
Appendix~\ref{appendix3d}).

The local emission (\ref{fl_emission}) is defined in nine points
of local energy $E_\loc$ that are equally spaced with the
central point at its maximum. The normalization constant
$N_0$ in the formula (\ref{emission}) is such that the total
integrated flux of the line is unity. The parameters defining
local emission in this model are (see Table~\ref{kyf1ll_par}):
\begin{description} \itemsep -2pt
 \item[{\tt PhoIndex}] -- photon index of primary power-law illumination,
 \item[{\tt height}] -- height above the black hole where the primary source
 is located for ${\tt height}>0$, and radial power-law index for
 ${\tt height}\le0$,
 \item[{\tt Erest}] -- rest energy of the line in keV,
 \item[{\tt sigma}] -- width of the line in eV.
\end{description}
If positive, the {\tt height} parameter works as a switch -- the exact
value present in the tables {\tt lamp.fits} must be chosen.
If the {\tt height} parameter is negative, then this model assumes that the
local emission is the same as in the {\sc kyg1line} model with the
parameters ${\tt alpha}=-{\tt height}$, ${\tt rb=0}$ and ${\tt limb}=-2$
({\tt PhoIndex} parameter is unused in this case).

\section{Compton reflection models}
We have developed two new relativistic continuum models -- lamp-post Compton
reflection model {\sc kyl1cr} and the {\sc kyh1refl} model which is a
relativistically blurred {\sc hrefl} model that is already present in
{\sc{}xspec}. Both of these models are non-axisymmetric.

\subsection{Non-axisymmetric lamp-post Compton reflection model
{\fontfamily{phv}\fontshape{sc}\selectfont kyl1cr}}
\label{kyl1cr}
The emission in this model is induced by the illumination of the disc from the
primary power-law source located on the axis at {\tt height} above the black
hole. As in every non-axisymmetric model the observed spectrum is computed
according to eq.~(\ref{emission}). The local emission is
\begin{eqnarray}
\label{cr_emission}
N_{\loc}(E_\loc) \hspace*{-0.5em} & = & \hspace*{-0.5em}
g_{\rm L}^{{\tt PhoIndex}-1}\frac{\sin\theta_{\rm L}
{\rm d}\theta_{\rm L}}{r\,{\rm d}r}\,\sqrt{1-\frac{2{\tt height}}
{{\tt height}^2+({\tt a/M})^2}}\,f(E_\loc;\mu_{\rm i},\mu_{\rm e})\quad
{\rm for}\quad {\tt height} > 0\, ,\hspace*{1em} \\
\label{cr_emission_neg}
N_{\loc}(E_\loc) \hspace*{-0.5em} & = & \hspace*{-0.5em} r^{\tt height}\,\bar{f}
(E_\loc;\mu_{\rm e})\quad {\rm for}\quad {\tt height \le 0}\, .
\end{eqnarray}
For the definition of $g_{\rm L}$, $\theta_{\rm L}$ and $\mu_{\rm i}$ see
Section \ref{section_kyf1ll} and Appendix~\ref{appendix3c}, where pre-calculated
tables of these functions in {\tt lamp.fits} are described.

The function $f(E_\loc;\mu_{\rm i},\mu_{\rm e})$ gives dependence of the locally
emitted spectrum on the angle of incidence and the angle of emission, assuming
a power-law illumination. This function depends on the photon index
{\tt PhoIndex} of the power-law emission from a primary source.
Values of this function for various photon indices of primary emission
are read from the pre-calculated tables stored in {\tt refspectra.fits}
(see Appendix~\ref{appendix3e}). These tables were calculated by the Monte Carlo
simulations of Compton scattering \cite{matt_1991}. At present,
tables for ${\tt PhoIndex}=1.5,\ 1.6,\ \dots,\ 2.9,\ 3.0$  and for local
 energies in the range from $2$~keV to $300$~keV are available.
The normalization constant $N_0$ in eq.~(\ref{emission}) is such
that the final photon flux at an energy of $3$~keV is equal to unity,
which is different from what is usual for
continuum models in {\sc xspec} (where the photon flux is unity at $1$~keV).
The choice adopted is due to the fact that current tables in
{\tt refspectra.fits} do not extend below $2$~keV.

The function $\bar{f}(E_\loc;\mu_{\rm e})$, which is used for negative
{\tt height}, is an averaged function $f(E_\loc;\mu_{\rm i},\mu_{\rm e})$ over
$\mu_{\rm i}$
\begin{equation}
\bar{f}(E_\loc;\mu_{\rm e})\equiv\int_0^1 {\rm d}\mu_{\rm i}\,f(E_\loc;
\mu_{\rm i},\mu_{\rm e})\, .
\end{equation}
The local emission (\ref{cr_emission_neg}) can be interpreted as emission
induced by illumination from clouds localized near above the disc rather than
from a primary source on the axis (see Figure~\ref{compton_reflection}). In this
case photons strike the disc from all directions.

For positive values of {\tt height} the {\sc kyl1cr} model includes a physical
model of polarization based on Rayleigh scattering in single scattering
approximation. The specific local Stokes parameters describing local
polarization of light are
\begin{eqnarray}
\label{polariz1}
i_\loc(E_\loc) & = & \frac{I_{\rm l}+I_{\rm r}}{\langle I_{\rm l}+
I_{\rm r}\rangle}\, N_\loc(E_\loc)\, ,\\[1mm]
q_\loc(E_\loc) & = & \frac{I_{\rm l}-I_{\rm r}}{\langle I_{\rm l}+
I_{\rm r}\rangle}\, N_\loc(E_\loc)\, ,\\[1mm]
u_\loc(E_\loc) & = & \frac{U}{\langle I_{\rm l}+
I_{\rm r}\rangle}\,N_\loc(E_\loc)\, ,\\[1mm]
\label{polariz4}
v_\loc(E_\loc) & = & 0\, ,
\end{eqnarray}
where functions $I_{\rm l}$, $I_{\rm r}$ and $U$ determine the angular
dependence of the Stokes parameters in the following way
\begin{eqnarray}
\label{polariz2}
\nonumber
I_{\rm l} & = & \mu_{\rm e}^2(1+\mu_{\rm i}^2)+2(1-\mu_{\rm e}^2)
(1-\mu_{\rm i}^2)-4\mu_{\rm e}\mu_{\rm i}\sqrt{(1-\mu_{\rm e}^2)
(1-\mu_{\rm i}^2)}\,\cos{(\Phi_{\rm e}-\Phi_{\rm i})}\\
 & & -\mu_{\rm e}^2(1-\mu_{\rm i}^2)\cos{2(\Phi_{\rm e}-\Phi_{\rm i})}\, ,\\
I_{\rm r} & = & 1+\mu_{\rm i}^2+(1-\mu_{\rm i}^2)\cos{2(\Phi_{\rm e}-
\Phi_{\rm i})}\, ,\\
U & = & -4\mu_{\rm i}\sqrt{(1-\mu_{\rm e}^2)(1-\mu_{\rm i}^2)}\,
\sin{(\Phi_{\rm e}-\Phi_{\rm i})}-2\mu_{\rm e}(1-\mu_{\rm i}^2)
\sin{2(\Phi_{\rm e}-\Phi_{\rm i})}\, ,
\end{eqnarray}
Here $\Phi_{\rm e}$ and $\Phi_{\rm i}$ are the azimuthal emission
and the incident angles in the local rest frame co-moving with the accretion
disc (see Appendixes~\ref{appendix1} and \ref{appendix2} for their
definition). For the derivation of these formulae see the definitions (I.147)
and eqs.~(X.172) in \cite{chandrasekhar_1960}.
We have omitted a common multiplication factor, which
would be cancelled anyway in eqs.~(\ref{polariz1})--(\ref{polariz4}).
The symbol $\langle\ \rangle$ in definitions of the local Stokes parameters
means value averaged over
the difference of the azimuthal angles $\Phi_{\rm e}-\Phi_{\rm i}$. We divide
the parameters by
$\langle I_{\rm l}+I_{\rm r}\rangle$ because the function
$f(E_\loc;\mu_{\rm i},\mu_{\rm e})$, and thus
also the local photon flux $N_\loc(E_\loc)$, is averaged over the difference of
the azimuthal angles.

\begin{table}[tbh]
\begin{center}
\begin{tabular}[h]{r@{}l|c|c|c|c}
\multicolumn{2}{c|}{parameter}  & unit & default value  & minimum value &
maximum value \\ \hline
&{\tt a/M}          & $GM/c$   & 0.9982   & 0.       & 1.        \\
&{\tt theta\_o}     & deg      & 30.      & 0.       & 89.       \\
&{\tt rin-rh}       & $GM/c^2$ & 0.       & 0.       & 999.      \\
&{\tt ms}           & --       & 1.       & 0.       & 1.        \\
&{\tt rout-rh}      & $GM/c^2$ & 400.     & 0.       & 999.      \\
&{\tt phi}          & deg      & 0.       & -180.    & 180.      \\
&{\tt dphi}         & deg      & 360.     & 0.       & 360.      \\
&{\tt nrad}         & --       & 200.     & 1.       & 10000.    \\
&{\tt division}     & --       & 1.       & 0.       & 1.        \\
&{\tt nphi}         & --       & 180.     & 1.       & 20000.    \\
&{\tt smooth}       & --       & 1.       & 0.       & 1.        \\
&{\tt zshift}       & --       & 0.       & -0.999   &  10.      \\
&{\tt ntable}       & --       & 0.       & 0.       & 99.       \\
{*}&{\tt PhoIndex}  & --       & 2.       & 1.5      & 3.        \\
{*}&{\tt height}    &$GM/c^2$  & 3.       & -20.     & 100.      \\
{*}&{\tt line}      & --       & 0.       & 0.       & 1.        \\
{*}&{\tt E\_cut}    & keV      & 300.     & 1.       & 1000.     \\
&{\tt Stokes}       & --       & 0.       & 0.       & 6.        \\
\end{tabular}
\end{center}
\caption{Parameters of the lamp-post Compton reflection model {\sc{}kyl1cr}.
Model parameters that are not common for all non-axisymmetric models are
denoted by asterisk.}
\label{kyl1cr_par}
\end{table}
The parameters defining local emission in this model are
(see Table~\ref{kyl1cr_par}):
\begin{description} \itemsep -2pt
 \item[{\tt PhoIndex}] -- photon index of primary power-law illumination,
 \item[{\tt height}] -- height above the black hole where the primary source
 is located for ${\tt height}>0$, and radial power-law index for
 ${\tt height}\le0$,
 \item[{\tt line}] -- switch whether to include the iron lines
 (0 -- no, 1 -- yes),
 \item[{\tt E\_cut}] -- exponential cut-off energy of the primary source in keV.
\end{description}
Tables {\tt refspectra.fits} for the function $f(E_\loc;\mu_{\rm i},\mu{\rm e})$
also contain the emission in the iron lines
K$\alpha$ and K$\beta$. The two lines can be excluded from
computations if the {\tt line} switch is set to zero.
The {\tt E\_cut} parameter sets the upper boundary in energies
where the emission from a primary source ceases to follow a power-law
dependence. If the {\tt E\_cut} parameter is lower than both the maximum energy
of the considered dataset and the maximum energy in the tables
for $f(E_\loc;\mu_{\rm i},\mu{\rm e})$ in {\tt refspectra.fits} (300~keV),
then this model is not valid.

\subsection{Non-axisymmetric Compton reflection model
{\fontfamily{phv}\fontshape{sc}\selectfont kyh1refl}}
This model is based on an existing multiplicative {\sc hrefl} model in
combination with the {\sc powerlaw} model, both of which are present in
{\sc xspec}. Local emission in (\ref{emission}) is the same as the spectrum
given by the model {\sc hrefl*powerlaw} with the parameters ${\tt thetamin}=0$
and ${\tt thetamax}=90$ with a broken power-law radial dependence added:
\begin{eqnarray}
N_\loc(E_\loc) & = & r^{-{\tt alpha}}\,\textsc{hrefl*powerlaw}\quad {\rm for}
\quad r\ge r_{\rm b}\, ,\\
N_\loc(E_\loc) & = & {\tt jump}\,r_{\rm b}^{{\tt beta}-{\tt alpha}}\,
r^{-{\tt beta}}\,\textsc{hrefl*powerlaw}\quad {\rm for} \quad r<r_{\rm b}\, .
\end{eqnarray}
For a definition of the boundary radius $r_{\rm b}$ by the {\tt rb} parameter
see eqs.~(\ref{rb1})--(\ref{rb2}), and
for a detailed description of the {\sc hrefl} model see \cite{dovciak04}
and the {\sc xspec} manual. The {\sc{}kyh1refl} model can be interpreted
as a Compton-reflection model for which the source of primary
irradiation is near above the disc, in contrast to the lamp-post
scheme with the source on the axis (see Figure~\ref{compton_reflection}).
The approximations for Compton reflection used in {\sc{}hrefl}
(and therefore also in {\sc{}kyh1refl})
are valid below $\sim15$~keV in the disc rest-frame.
The normalization of the final spectrum in this model is the same as
in other continuum models in {\sc xspec}, i.e.\ photon flux is unity at
the energy of $1$~keV.

\begin{table}[tbh]
\begin{center}
\begin{tabular}[h]{r@{}l|c|c|c|c}
\multicolumn{2}{c|}{parameter} & unit & default value  & minimum value &
maximum value \\ \hline
&{\tt a/M}         & $GM/c$   & 0.9982 & 0.      & 1.      \\
&{\tt theta\_o}    & deg      & 30.    & 0.      & 89.     \\
&{\tt rin-rh}      & $GM/c^2$ & 0.     & 0.      & 999.    \\
&{\tt ms}          & --       & 1.     & 0.      & 1.      \\
&{\tt rout-rh}     & $GM/c^2$ & 400.   & 0.      & 999.    \\
&{\tt phi}         & deg      & 0.     & -180.   & 180.    \\
&{\tt dphi}        & deg      & 360.   & 0.      & 360.    \\
&{\tt nrad}        & --       & 200.   & 1.      & 10000.  \\
&{\tt division}    & --       & 1.     & 0.      & 1.      \\
&{\tt nphi}        & --       & 180.   & 1.      & 20000.  \\
&{\tt smooth}      & --       & 1.     & 0.      & 1.      \\
&{\tt zshift}      & --       & 0.     & -0.999  & 10.     \\
&{\tt ntable}      & --       & 0.     & 0.      & 99.     \\
{*}&{\tt PhoIndex} & --       & 1.     & 0.      & 10.     \\
{*}&{\tt alpha}    & --       & 3.     & -20.    & 20.     \\
{*}&{\tt beta}     & --       & 4.     & -20.    & 20.     \\
{*}&{\tt rb}       & $r_{\rm ms}$ & 0. & 0.      & 160.    \\
{*}&{\tt jump}     & --       & 1.     & 0.      & 1e6     \\
{*}&{\tt Feabun}   & --       & 1.     & 0.      & 200.    \\
{*}&{\tt FeKedge}  & keV      & 7.11   & 7.0     & 10.     \\
{*}&{\tt Escfrac}  & --       & 1.     & 0.      & 1000.   \\
{*}&{\tt covfac}   & --       & 1.     & 0.      & 1000.   \\
&{\tt Stokes}      & --       & 0.     & 0.      & 6.      \\
\end{tabular}
\end{center}
\caption{Parameters of the reflection {\sc{}kyh1refl} model. Model parameters
that are not common for all non-axisymmetric models are denoted by asterisk.}
\label{kyh1refl_par}
\end{table}
The parameters defining the local emission in {\sc kyh1refl}
(see Table~\ref{kyh1refl_par}) are
\begin{description} \itemsep -2pt
 \item[{\tt PhoIndex}] -- photon index of the primary power-law illumination,
 \item[{\tt alpha}] -- radial power-law index for the outer region,
 \item[{\tt beta}] -- radial power-law index for the inner region,
 \item[{\tt rb}] -- parameter defining the border between regions with different
 power-law indices,
 \item[{\tt jump}] -- ratio between flux in the inner and outer regions at
 the border radius,
 \item[{\tt Feabun}] -- iron abundance relative to solar,
 \item[{\tt FeKedge}] -- iron K-edge energy,
 \item[{\tt Escfrac}] -- fraction of the direct flux from the power-law primary
 source seen by the observer,
 \item[{\tt covfac}] -- normalization of the reflected continuum.
\end{description}

\section{General relativistic convolution models}
We have also produced two convolution-type
models, {\sc ky1conv} and {\sc kyconv}, which can be applied to any existing
{\sc{}xspec} model for the intrinsic X-ray emission from a disc around a Kerr
black hole. We must stress
that these models are substantially more powerful than the usual
convolution models in {\sc{}xspec} (these are commonly
defined in terms of one-dimensional integration over energy bins).
Despite the fact that our convolution models still use the standard
{\sc{}xspec} syntax in evaluating the observed spectrum
(e.g.\ {\sc kyconv(powerlaw)}), our code
accomplishes a more complex operation. It still performs ray-tracing
across the disc surface so that the intrinsic model contributions are
integrated from different radii and azimuths on the disc.

There are several restrictions that arise from the fact that we use existing
{\sc xspec} models:
\begin{itemize} \itemsep -2pt
 \item[--] by local {\sc xspec} models only the energy dependence of the photon
 flux can be defined,
 \item[--] only a certain type of radial dependence of the local photon flux can
 be imposed -- we have chosen to use a broken power-law radial dependence,
 \item[--] there is no azimuthal dependence of the local photon flux, except
 through limb darkening law,
 \item[--] local flux depends on the binning of the data because it is defined
 in the centre of each bin, a large number of bins is needed for highly varying
 local  flux.
\end{itemize}

For emissivities that cannot be defined by existing {\sc xspec} models,
or where the limitations mentioned above are too restrictive, one has
to add a new user-defined model to {\sc{}xspec} (by adding a new
subroutine to {\sc xspec}). This method is more flexible and faster than
convolution models (especially when compared with the non-axisymmetric
one), and hence it is recommended even for cases when these
prefabricated models could be used. In any new model for {\sc
xspec} one can use the common ray-tracing driver for relativistic smearing
of the local emission: {\tt ide} for non-axisymmetric
models and {\tt idre} for axisymmetric ones. For a detailed description
see Appendixes~\ref{appendix4a} and \ref{appendix4b}.

\subsection{Non-axisymmetric convolution model
{\fontfamily{phv}\fontshape{sc}\selectfont kyc1onv}}
The local emission in this model is computed according to
eq.~(\ref{emission}) with the local emissivity equal to
\begin{eqnarray}
N_\loc(E_\loc) & = & r^{-{\tt alpha}}\,f(\mu_{\rm e})\,\textsc{model}\quad
{\rm for}\quad r > r_{\rm b} \, ,\\
N_\loc(E_\loc) & = & {\tt jump}\,r_{\rm b}^{{\tt beta}-{\tt alpha}}\,
r^{-{\tt beta}}\,f(\mu_{\rm e})\,\textsc{model} \quad {\rm for}\quad r
\le r_{\rm b} \, .\
\end{eqnarray}
For a definition of the boundary radius $r_{\rm b}$ by the {\tt rb} parameter
see eqs.~(\ref{rb1})--(\ref{rb2}) and for definition of different limb
darkening laws $f(\mu_{\rm e})$ see eqs.~(\ref{isotropic})--(\ref{other_limb}).
The local emission is given by the
{\sc model} in the centre of energy bins used in {\sc xspec} with the
broken power-law radial dependence and limb darkening law added. Apart from the
parameters of the {\sc model}, the local emission is defined also by the
following parameters (see Table~\ref{kyc1onv_par}):
\begin{description} \itemsep -2pt
 \item[{\tt normal}] -- switch for the normalization of the final spectrum,\\
   $=$ 0 -- total flux is unity (used usually for the line),\\
   $>$ 0 -- flux is unity at the energy = {\tt normal} keV (used usually for
   the continuum),\\
   $<$ 0 -- flux is not normalized,
 \item[{\tt ne\_loc}] -- number of points in the energy grid where the local
 photon flux is defined,
 \item[{\tt alpha}] --  radial power-law index for the outer region,
 \item[{\tt beta}] -- radial power-law index for the inner region,
 \item[{\tt rb}] -- parameter defining the border between regions with different
 power-law indices,
 \item[{\tt jump}] -- ratio between the flux in the inner and outer regions at
 the border radius,
 \item[{\tt limb}] -- switch for different limb darkening/brightening laws.
\end{description}
The local emission in each {\sc ky} model has to be defined either on
equidistant or exponential (i.e.\ equidistant in logarithmic scale)
energy grid. Because the energy grid used in the convolution model depends on
the binning of the data, which may be arbitrary, the flux has to be
rebinned. It is always rebinned into an exponentially spaced
energy grid in {\sc ky} convolution models.
The {\tt ne\_loc} parameter defines the number of points in which the rebinned
flux will be defined.

\begin{table}[tbh]
\begin{center}
\begin{tabular}[h]{r@{}l|c|c|c|c}
\multicolumn{2}{c|}{parameter} & unit & default value  & minimum value &
maximum value \\ \hline
&{\tt a/M}         & $GM/c$   & 0.9982   & 0.       & 1.        \\
&{\tt theta\_o}    & deg      & 30.      & 0.       & 89.       \\
&{\tt rin-rh}      & $GM/c^2$ & 0.       & 0.       & 999.      \\
&{\tt ms}          & --       & 1.       & 0.       & 1.        \\
&{\tt rout-rh}     & $GM/c^2$ & 400.     & 0.       & 999.      \\
&{\tt phi}         & deg      & 0.       & -180.    & 180.      \\
&{\tt dphi}        & deg      & 360.     & 0.       & 360.      \\
&{\tt nrad}        & --       & 200.     & 1.       & 10000.    \\
&{\tt division}    & --       & 1.       & 0.       & 1.        \\
&{\tt nphi}        & --       & 180.     & 1.       & 20000.    \\
&{\tt smooth}      & --       & 1.       & 0.       & 1.        \\
{*}&{\tt normal}   & --       & 1.       & -1.      & 100.      \\
&{\tt zshift}      & --       & 0.       & -0.999   & 10.       \\
&{\tt ntable}      & --       & 0.       & 0.       & 99.       \\
{*}&{\tt ne\_loc}  & --       & 100.     & 3.       & 5000.     \\
{*}&{\tt alpha}    & --       & 3.       & -20.     & 20.       \\
{*}&{\tt beta}     & --       & 4.       & -20.     & 20.       \\
{*}&{\tt rb}       & $r_{\rm ms}$ & 0.   & 0.       & 160.      \\
{*}&{\tt jump}     & --       & 1.       & 0.       & 1e6       \\
{*}&{\tt limb}     & --       & 0.       & -10.     & 10.       \\
&{\tt Stokes}      & --       & 0.       & 0.       & 6.        \\
\end{tabular}
\end{center}
\caption{Parameters of the non-axisymmetric convolution model {\sc{}kyc1onv}.
Model parameters that are not common for all non-axisymmetric models are
denoted by asterisk.}
\label{kyc1onv_par}
\end{table}

\subsection{Axisymmetric convolution model
{\fontfamily{phv}\fontshape{sc}\selectfont kyconv}}
The local emission in this model is computed according to
eq.~(\ref{axisym_emission}) with the local emissivity equal to
\begin{eqnarray}
N_\loc(E_\loc) & = & \textsc{model}\, ,\\
R(r) & = & r^{-{\tt alpha}}\, .\\
\end{eqnarray}
Except for the parameters of the {\sc model}, the local emission is defined also
by the following parameters (see Table~\ref{kyconv_par}):
\begin{description} \itemsep -2pt
 \item[{\tt alpha}] --  radial power-law index,
 \item[{\tt ne\_loc}] -- number of points in energy grid where the local photon
 flux is defined,
 \item[{\tt normal}] -- switch for the normalization of the final spectrum,\\
   $=$ 0 -- total flux is unity (used usually for the line),\\
   $>$ 0 -- flux is unity at the energy = {\tt normal} keV (used usually for
   the continuum),\\
   $<$ 0 -- flux is not normalized.
\end{description}
Note that the limb darkening/brightening law can be chosen through the {\tt ntable}
switch. This model is much faster than the non-axisymmetric convolution model
{\sc kyc1onv}.

\begin{table}[tbh]
\begin{center}
\begin{tabular}[h]{r@{}l|c|c|c|c}
\multicolumn{2}{c|}{parameter} & unit & default value  & minimum value &
maximum value \\ \hline
&{\tt a/M}         & $GM/c$   & 0.9982   & 0.       & 1.        \\
&{\tt theta\_o}    & deg      & 30.      & 0.       & 89.       \\
&{\tt rin-rh}      & $GM/c^2$ & 0.       & 0.       & 999.      \\
&{\tt ms}          & --       & 1.       & 0.       & 1.        \\
&{\tt rout-rh}     & $GM/c^2$ & 400.     & 0.       & 999.      \\
&{\tt zshift}      & --       & 0.       & -0.999   & 10.       \\
&{\tt ntable}      & --       & 0.       & 0.       & 99.       \\
{*}&{\tt alpha}    & --       & 3.       & -20.     & 20.       \\
{*}&{\tt ne\_loc}  & --       & 100.     & 3.       & 5000.     \\
{*}&{\tt normal}   & --       & 1.       & -1.      & 100.      \\
\end{tabular}
\end{center}
\caption{Parameters of the axisymmetric convolution model {\sc{}kyconv}.
Model parameters that are not common for all axisymmetric models are denoted
by asterisk.}
\label{kyconv_par}
\end{table}

\section{Examples and comparisons}

\begin{figure*}[tb]
  \begin{center}
  \begin{tabular}{cc}
   \hspace*{-2.5mm}\includegraphics[width=6.75cm]{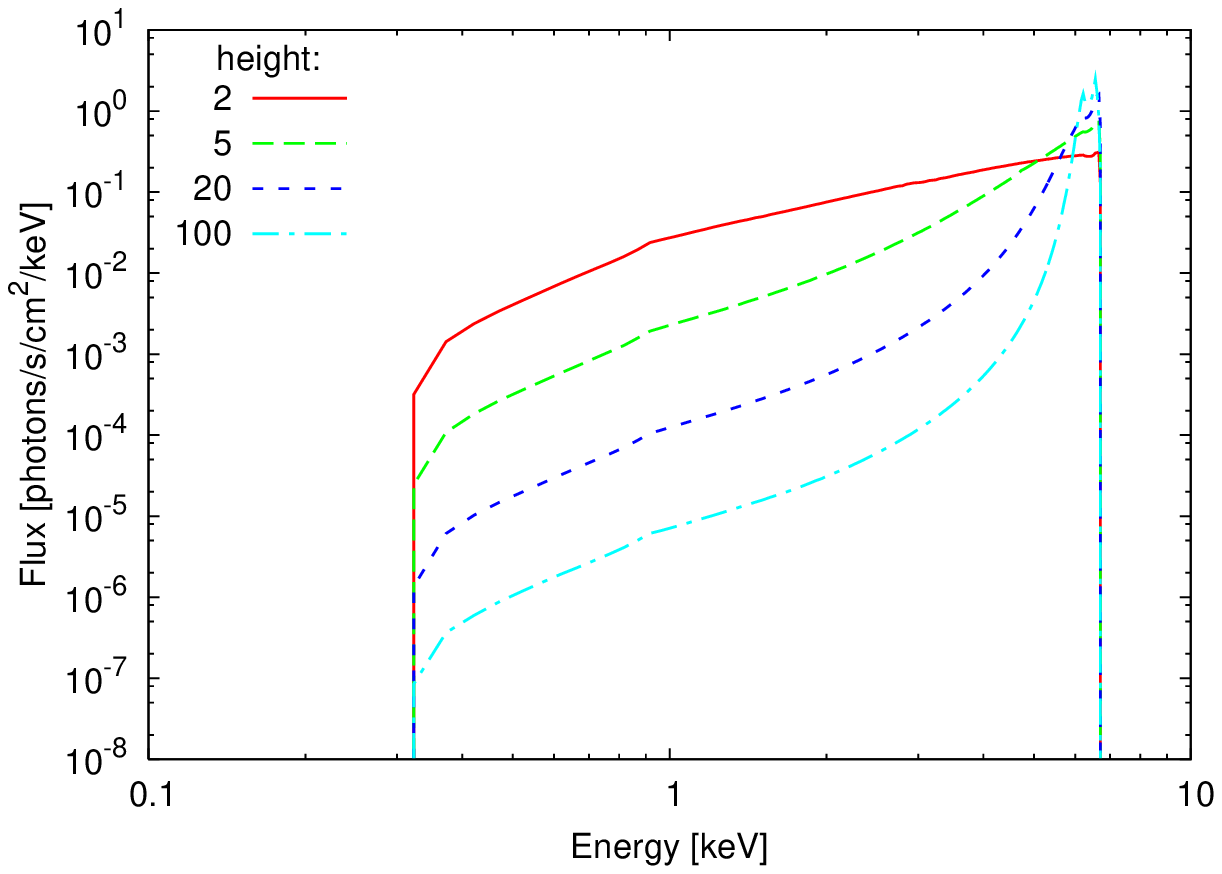} &
   \hspace*{-2.5mm}\includegraphics[width=6.75cm]{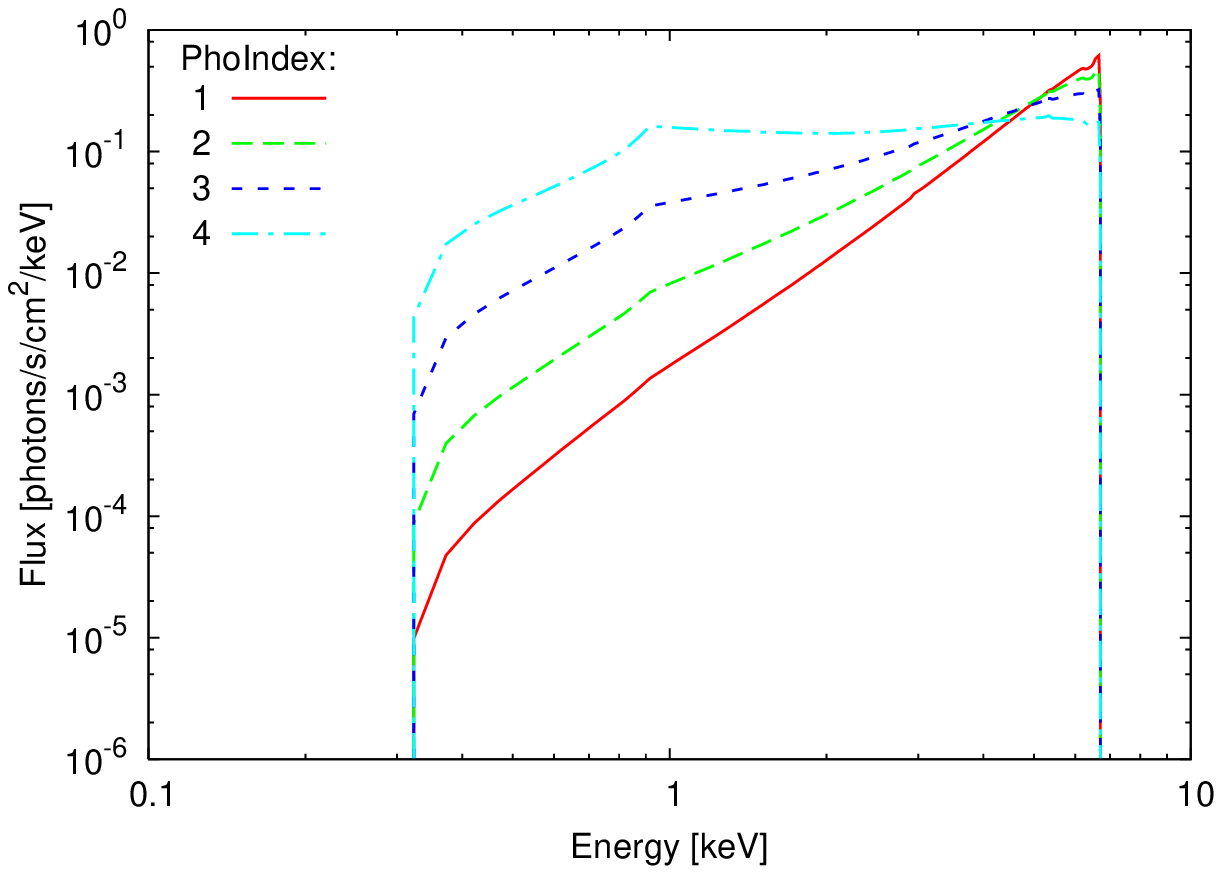}
  \end{tabular}
 \end{center}
\caption{An example of a line profile originating from a disc in
equatorial plane of a Kerr black hole ($a=0.9987\,GM/c^2$,
i.e.\ $r_{\rm h}=1.05\,GM/c^2$) due to the illumination
from a primary source on the axis. The {\sc kyf1ll} model was used. Left: Dependence
of the line profile on the
height (in $GM/c^2$) of a primary source with photon index $\Gamma=2$.
Right: Dependence of the line profile on the photon index of the primary
emission with a source at height $3\,GM/c^2$ above the black hole.}
\label{kyf}
\end{figure*}

\begin{figure*}[tb]
 \begin{center}
  \begin{tabular}{cc}
   \includegraphics[width=6.5cm]{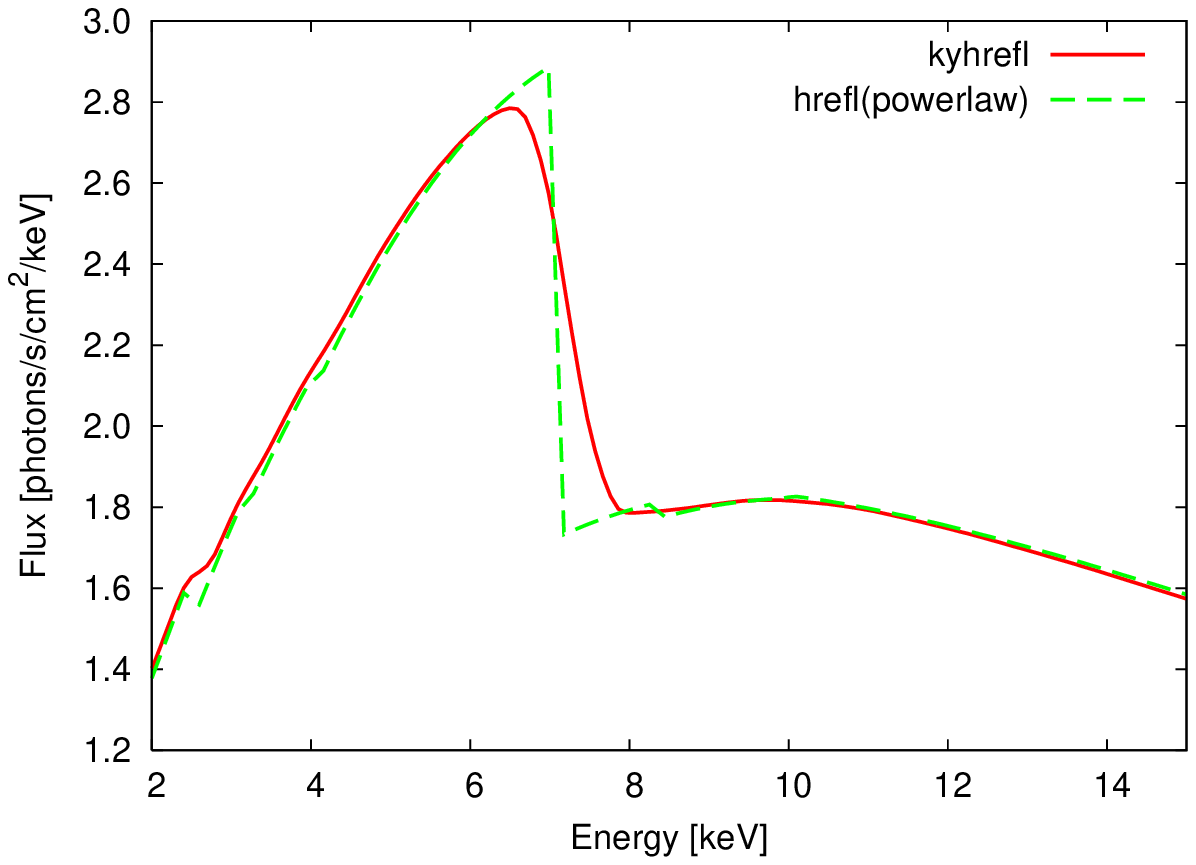} &
   \includegraphics[width=6.5cm]{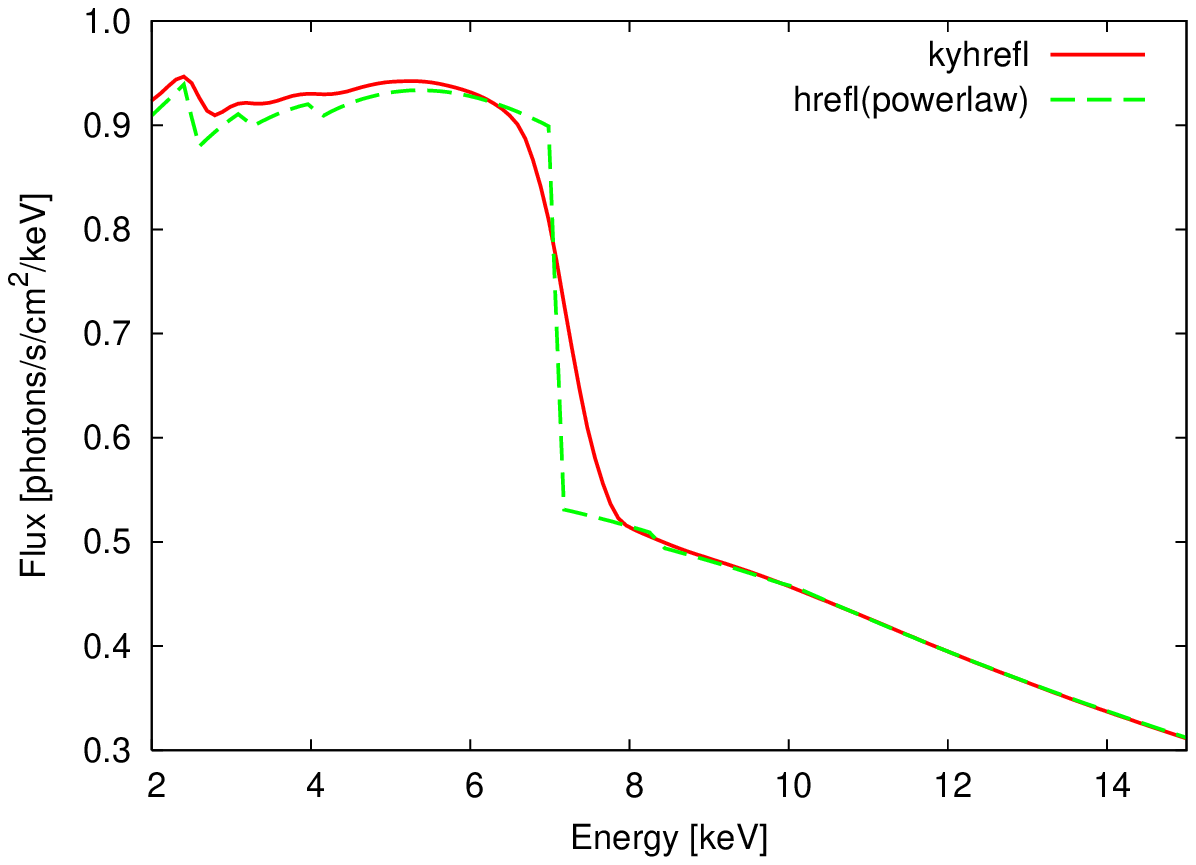}
  \end{tabular}
 \end{center}
 \caption{Comparison of the general relativistic {\sc kyh1refl} model with
 non-relativistic {\sc hrefl(powerlaw)}. The relativistic blurring of the iron
 edge is clearly visible. The power-law index of the primary source is
 {\tt PhoIndex}=2 (left) and {\tt PhoIndex}=2.6 (right).}
 \label{kyh_href}
\end{figure*}

\begin{figure*}[tb]
 \begin{center}
  \begin{tabular}{cc}
   \includegraphics[width=6.5cm]{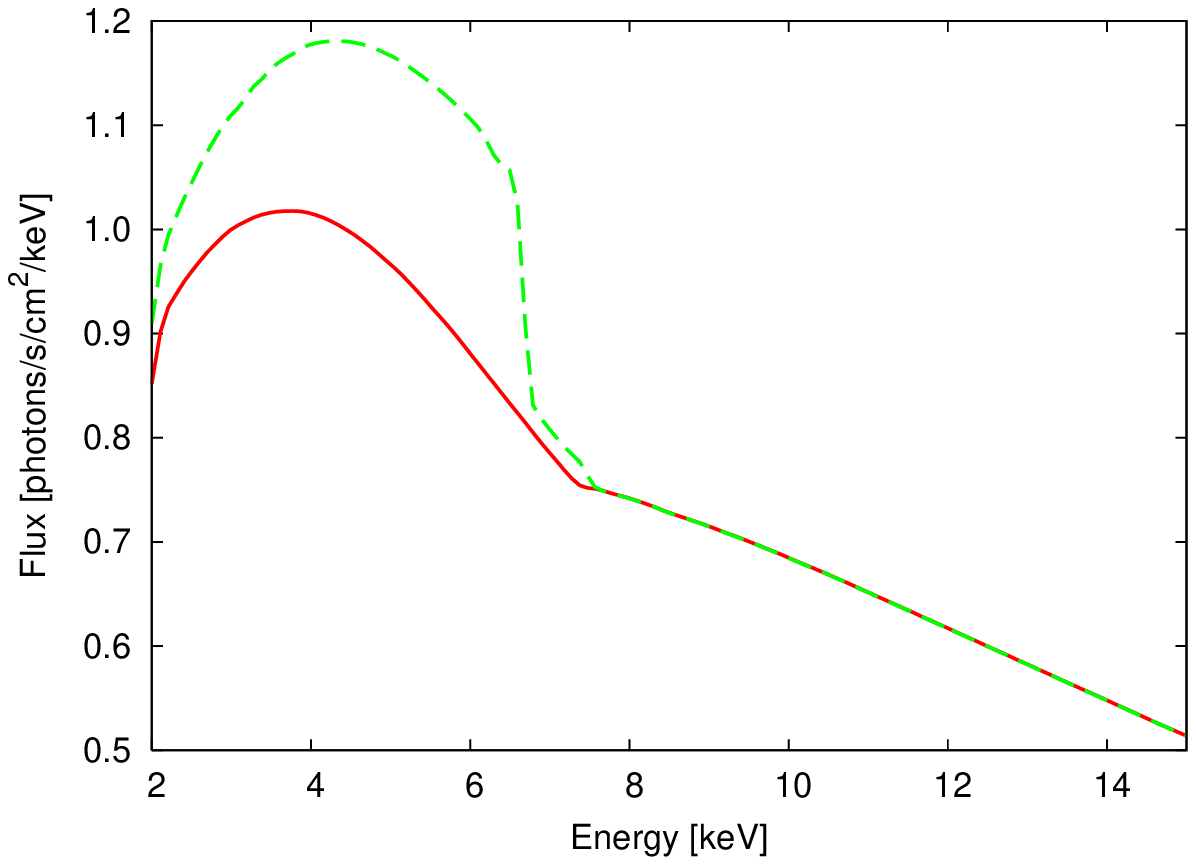} &
   \includegraphics[width=6.5cm]{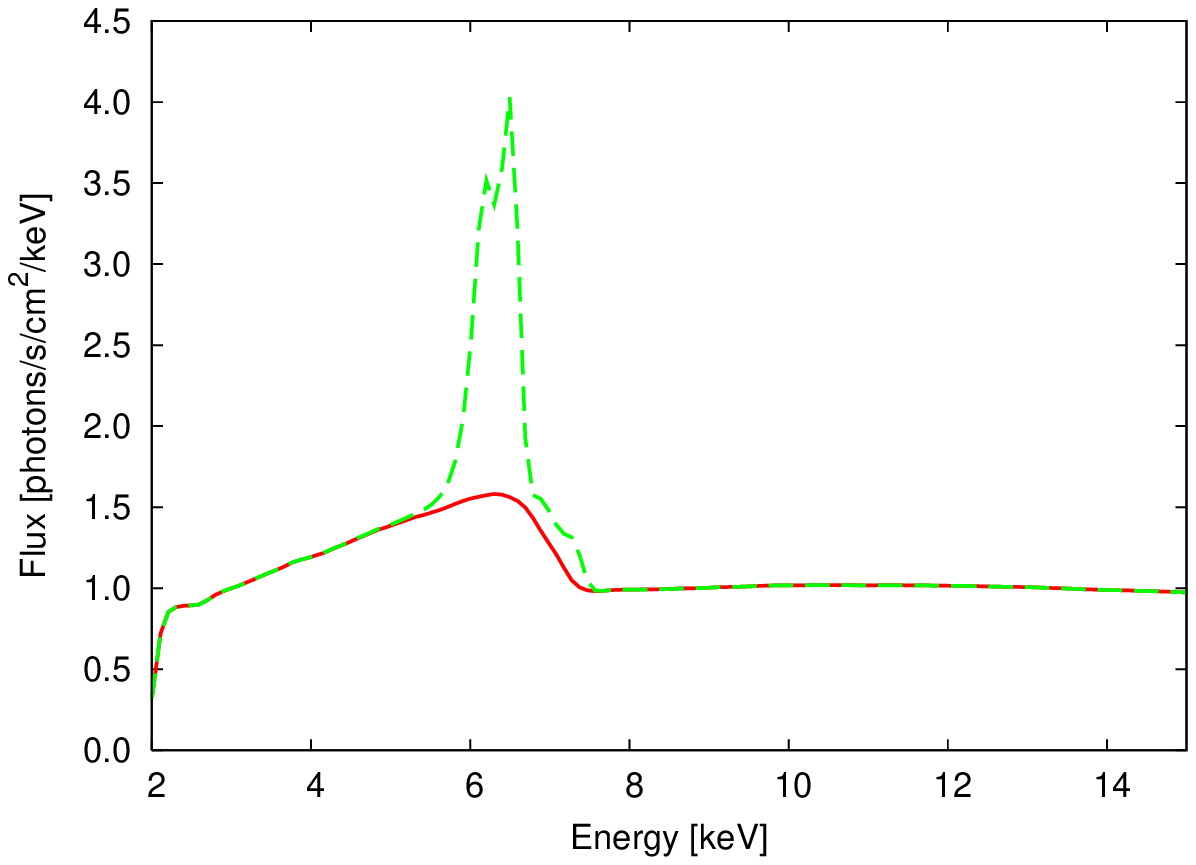}
  \end{tabular}
 \end{center}
 \caption{General relativistic lamp-post Compton reflection model {\sc kyl1cr}
 with (dashed) and without (solid) iron lines K$\alpha$ and K$\beta$. The emission
 from the disc is induced by illumination from a primary source placed
 $2\,GM/c^2$ (left) and $100\,GM/c^2$ (right) above the black hole.}
 \label{kyl_kyl}
\end{figure*}

\begin{figure*}[tb]
 \begin{center}
  \begin{tabular}{cc}
   \includegraphics[width=6.5cm]{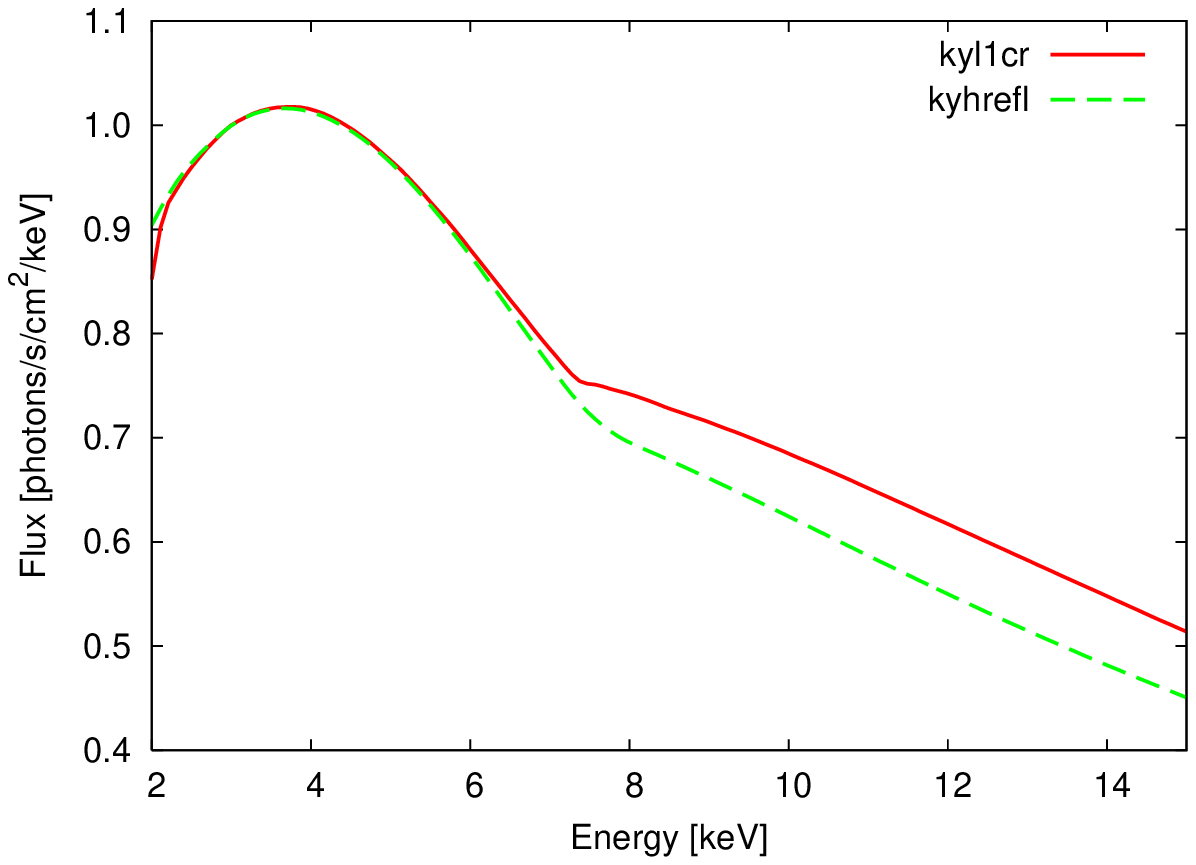} &
   \includegraphics[width=6.5cm]{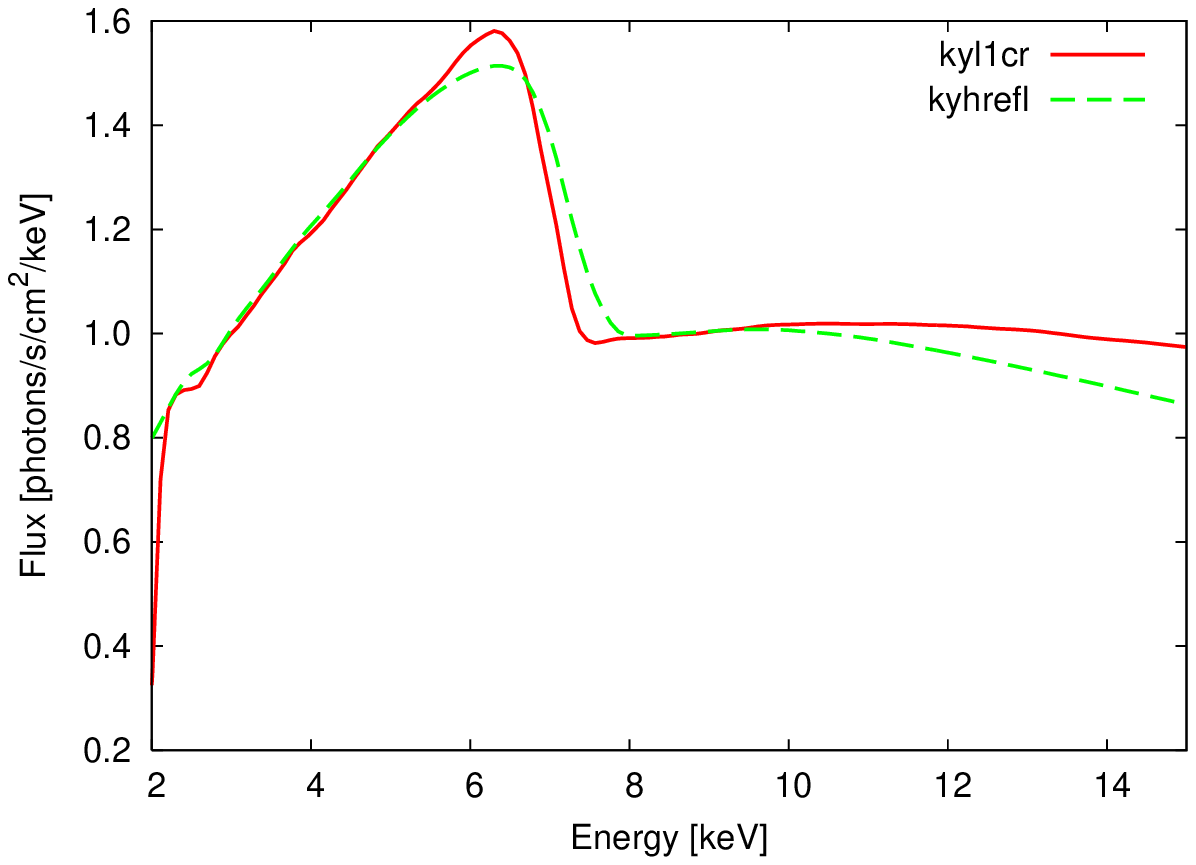}
  \end{tabular}
 \end{center}
 \caption{Comparison of the two new general relativistic Compton reflection
 models {\sc kyl1cr} and
 {\sc kyh1refl}. The lamp-post {\sc kyl1cr} model is characterized by the height
 $h$ above the disc where a primary source of emission is placed, the
 reflection {\sc kyh1refl} model is characterized by the radial power-law index
 $\alpha$. Left: $h=2\,GM/c^2$, $\alpha=3.4\,$. Right:
 $h=100\,GM/c^2$, $\alpha=1.5\,$.}
 \label{kyl_kyh}
\end{figure*}

In our new models we have concentrated ourselves mainly on two components
that contribute to the X-ray spectra of active galactic nuclei and X-ray
binaries with black-hole candidates -- spectral line emission and its
relativistic broadening,
and the Compton reflection from an illuminated disc. Two basic types of
illumination have been considered -- the disc illuminated either from
every direction by a nearby diffuse corona
above the disc, or from a particular direction by a small source placed on the
axis above the black hole (see Figure~\ref{compton_reflection}).
The illumination in the former case decreases with radius as a power law.
Hence, this model is characterized by the radial
power-law index $\alpha$. On the other hand, the illumination anisotropy
in the latter (lamp-post)
model depends mainly on the position of a primary source of emission
characterized by height $h$ where it is located. In both cases it is assumed
that the primary emission has a continuum power-law shape which can be
characterized by a photon index $\Gamma$ ({\tt PhoIndex}).

The emission from the disc depends on quite a number of parameters. It is
influenced by the mass $M$ and the rotation $a$ of the central black hole,
by the area from which the emission from disc comes (defined by inner radius
$r_{\rm in}$, outer radius $r_{\rm out}$ and azimuthal segment with boundaries
at $\varphi$ and
$\varphi+\Delta\varphi$), by the inclination $\theta_{\rm o}$ of the observer,
by the radial power-law index $\alpha$, by the photon index $\Gamma$ and
by the position $h$ of a primary source.
Limb darkening/brightening law (dependence on the local emission angle)
is another important factor that determines the final spectrum we observe.

Here, we will show several examples of emission for the lamp-post fluorescent line
model and for the reflection models. For other examples and comparisons, see the
accompanying paper (Dov\v{c}iak, Karas \& Yaqoob \cite{dovciak04}). In all
figures in this section, we assumed the inclination angle
$\theta_{\rm o}=30^\circ$, the rotational parameter $a=0.9987\,GM/c^2$, and an
emitting ring extending from $r_{\rm in}=r_{\rm ms}$ to $r_{\rm out}=400\,GM/c^2$.

In Figure~\ref{kyf} we demonstrate that the broad iron emission lines due to
illumination from the source placed on the axis depend heavily on the height
where the ``lamp'' is located (left), as well as on the photon index of the
primary emission (right). It can be seen that the intrinsic width of
the line ($2\,{\rm eV}$ in this example) is much less than its subsequent
relativistic broadening, and the local profile (assumed to be Gaussian)
is thus smeared in the  final spectrum. These graphs correspond to the
iron K$\alpha$ line with the rest energy of $6.4$~keV.

Relativistic effects are demonstrated also in Figure~\ref{kyh_href} where
the non-relativistic reflection model {\sc hrefl(powerlaw)} is compared with
our relativistic {\sc kyh1refl}. Blurring of the iron edge is clearly
visible. Here, we set the radial power-law index $\alpha=1$ in
{\sc kyh1refl}. Other parameters defining these models were set to their default
values.

Examples of the Compton reflection emission component of the spectra with and
without the
fluorescent K$\alpha$ and K$\beta$ lines are shown in Figure~\ref{kyl_kyl}.
It can be seen that originally narrow lines can contribute substantially to the
continuum component.

We compare the two new relativistic reflection models {\sc kyl1cr} and
{\sc kyh1refl} in Figure~\ref{kyl_kyh}. Note that the {\sc kyl1cr} model is
valid only above approximately $2\,{\rm keV}$ and the {\sc kyh1refl} model only
below approximately $15\,{\rm keV}$.

\section{Conclusions}
In this paper we described the main features of the newly developed
set of routines. We have concentrated ourselves on various technical
issues connected with fitting X-ray spectra using our model.
In particular, we described
several variants of the code which are suited for modelling
relativistic spectral components originating in a Keplerian disc
near a rotating black hole.
Both axially symmetric and non-axisymmetric models were discussed.
For further details and for exemplary analysis of {\it XMM-Newton}
satellite data we refer to Dov\v{c}iak et al.\ \cite{dovciak04}.

Our package offers a number of applications which could not be examined
in the limited space of the present paper. In particular, timing
analysis can be performed with the code, but we defer detailed description
of this capability to subsequent papers. Also, we have only briefly
touched the possibility of polarimetric analysis, which offers great
possibilities for future X-ray spectroscopy but goes beyond routine
capabilities of devices installed onboard present day satellites.
Additional emissivity laws can be easily adopted. This can be
achieved either by using the convolution component or by
adding a new user-defined model. The latter method is
more flexible and faster, and hence recommended. In both approaches, the
ray-tracing routine is linked and used for relativistic blurring.

As general motivation for developing this project further, we remind the
reader that various disc-like structures are almost ubiquitous in objects
where the fluid orbits around and inflows onto a compact body.
The central mass, $\mbh$, can vary by many orders
of magnitude in different objects, and its value provides the basic
classification for black-hole sources. Physical characteristics of
accretion discs also scale roughly with $\mbh$. Indeed, accretion discs
around supermassive black holes in active galactic nuclei and
quasars share some properties with circumstellar discs in close
binary systems, e.g.\ cataclysmic variable stars and
microquasars. However, there are important distinctions between
the two kinds of objects which prohibit any simple scaling (for example,
galactic nuclear discs tend to be colder and less dense compared to
circumstellar discs). In both cases there is strong evidence
suggesting that some spectral components (namely, the iron K$\alpha$
line emission) originate, at least in part, within $\sim10$
gravitational radii of a central black-hole.

The central compact body governs gravitational field in which the medium
of an accretion flow evolves. Since we
consider a general relativistic description of the gravitational field,
the rotation of the central body should not be ignored. An angular momentum is
actually one of the model parameters which could in principle be
measured by means of spectral analysis of observed radiation.

\medskip

The authors gratefully acknowledge kind invitation of the meeting
organizers and financial support via the Czech Science Foundation
grants 205/03/0902 (VK), 202/02/0735 (MD), and from the NASA grants
NCC5-447 and NAG5-10769 (TY). Support from the Charles University
is also acknowledged (GAUK 299/2004). AM acknowledges financial support
from CNES and kind hospitality at the Astronomical Institute of the
Czech Academy of Sciences.

\appendix
\renewcommand{\theequation}{A\arabic{equation}}
\section{Appendix}
\subsection{Summary of equations}
\label{appendix1}
Before writing equations for the transfer functions let us summarize
basic formulae defining the Kerr space-time, light geodesics and 
disc's motion. We remind the reader that units $G\mbh=c=1$ are used
($\mbh$ is the mass of the central black hole).

The Kerr metric in Boyer-Lindquist coordinates is
$$g_{\mu\nu} = \left(
\begin{array}{cccc}
-(1-\frac{2r}{\rho^2}) & 0 & 0 & -\frac{2ar\sin^{2}
{\! \theta}}{\rho^2} \\[2mm]
0 & \frac{\rho^2}{\Delta} & 0 & 0 \\[2mm] 0 & 0 & \rho^2 & 0 \\[2mm]
-\frac{2ar\sin^{2}{\!\theta}}{\rho^2} & 0 & 0 & \frac{{\cal A}
\sin^{2}{\!\theta}}{\rho^2}
\end{array}
\right)\, ,$$ \\[2mm]
where $\rho^2 \equiv r^2+a^2\cos^2\!\theta$, $\Delta \equiv r^2-2r+a^2$
and ${\cal A} \equiv (r^2+a^2)^2-\Delta a^2\sin^2\!\theta$. We assume
$0\leq a\leq 1$ everywhere in this paper.

The four-momentum of the photons emitted from the disc is (see e.g.\
\cite{carter_1968} and \cite{misner_1973})
\begin{eqnarray}
p_{\rm e}^t & = & [a(l-a)+(r^2+a^2)(r^2+a^2-al)/\Delta]/r^2\, ,\\
p_{\rm e}^r & = & {\rm R}_{\rm sgn}\{(r^2+a^2-al)^2-
\Delta[(l-a)^2+q^2]\}^{1/2}/r^2\, ,\\
p_{\rm e}^{\theta} & = & -q/r^2\, ,\\
p_{\rm e}^{\varphi} & = & [l-a+a(r^2+a^2-al)/\Delta]/r^2\, .
\end{eqnarray}
Here $l=\alpha\sin{\theta_{\rm o}}$ and $q^2=\beta^2+\cos^2{\!(\alpha^2-a^2)}$
are constants of motion with $\alpha$ and $\beta$ being impact parameters
measured perpendicular and parallel, respectively, to the spin axis of the
black hole projected onto the observer's sky. Here we define $\alpha$ to
be positive when photon travels in the direction of the four-vector
$\frac{\partial}{\partial \varphi}$ at infinity, and $\beta$ to be positive if it
travels in the direction of $-\frac{\partial}{\partial\theta}$ at infinity.
Furthermore, we have denoted sign of the radial component of the momentum
by ${\rm R}_{\rm sgn}$. We have chosen an
affine parameter along light geodesics in such a way that the
conserved energy is normalized to $-p_{{\rm e}\,t}=-p_{{\rm o}\,t}=1$.

The Keplerian velocity of the co-rotating disc above the marginally stable orbit
is
\begin{eqnarray}
U^t & = & \frac{r^2+a\sqrt{r}}{r\sqrt{r^2-3r+2a\sqrt{r}}}\, ,\\
U^r & = & 0\, ,\\
U^\theta & = & 0\, ,\\
U^{\varphi} & = & \frac{1}{\sqrt{r(r^2-3r+2a\sqrt{r})}}\, .
\end{eqnarray}
We assume that the matter in the disc below the marginally stable orbit
conserves its specific energy and its specific angular momentum, i.e.\
$U_t(r<r_{\rm ms})=U_t(r_{\rm ms})$ and $U_\varphi(r<r_{\rm ms})=
U_\varphi(r_{\rm ms})$. We get the radial component $U^r(r<r_{\rm ms})$ from
the normalization of the four-velocity, $U^\mu\,U_\mu=-1$, and from the fact
that the disc rotates in the equatorial plane even below the marginally stable
orbit, i.e.\ $U^\theta(r<r_{\rm ms})=0$.

In our calculations we use the following local orthonormal tetrad connected
with the matter in the disc
\begin{eqnarray}
e_{(t)\,\mu} & = & U_\mu\, ,\\
e_{(r)\,\mu} & = & \left ( \frac{\sqrt{g_{rr}}\,U^rU_t}{\sqrt{1+U^rU_r}},\,
\sqrt{g_{rr}(1+U^rU_r)},\,0,\,\frac{\sqrt{g_{rr}}\,U^rU_\varphi}
{\sqrt{1+U^rU_r}}\right )\, ,\\
e_{(\theta)\,\mu} & = & (0,\,0,\,\sqrt{g_{\theta\theta}},\,0)\, ,\\
e_{(\varphi)\,\mu} & = & \left ( -\frac{\sqrt{\Delta}\,U^\varphi}
{\sqrt{1+U^rU_r}},\,0,\,0,\,\frac{\sqrt{\Delta}\,U^t}{\sqrt{1+U^rU_r}}\right )
\, .
\end{eqnarray}

The gravitational and Doppler shift ($g$-factor) is defined as the ratio of the
energy of a photon received by an observer at infinity
to the local energy of the same photon when emitted from an accretion disc
\begin{equation}
\label{gfac}
g=\frac{\nu_{\rm o}}{\nu_{\rm e}}=\frac{{p_{{\rm o}\,t}}}{{p_{{\rm e}\,\mu}}\,
U^{\mu}}=-\frac{1}{{p_{{\rm e}\,\mu}}\,
U^{\mu}}\, .
\end{equation}
Here $\nu_{\rm o}$ and $\nu_{\rm e}$ denote frequency of the observed and
emitted photons, respectively.

We define lensing as the ratio of the area at infinity perpendicular to the
light rays through which photons come to the proper area on the disc
perpendicular to the light rays and corresponding to the same flux tube
\begin{equation}
\label{lensing}
\frac{{\rm d}S_{\rm f}}{{\rm d}S_\perp}=\frac{1}{\sqrt{\|Y_{\rm e1}\|^2
\|Y_{\rm e2}\|^2 - <Y_{\rm e1},Y_{\rm e2}>^2}}\, .
\end{equation}
The four-vectors $Y_{\rm e1}$ and $Y_{\rm e2}$ are transported along the
geodesic according to the equation of the geodesic deviation from infinity
where they are unit, space-like and perpendicular to each other and to the
four-momentum of light.
In (\ref{lensing}) we have denoted the magnitude of a four-vector by $\|\ \|$
and the scalar product of two four-vectors by $<\ ,\ >$.

The cosine of the local emission angle is
\begin{equation}
\label{cosine}
\mu_{\rm e}=\cos{\,\delta_{\rm e}}=
-\displaystyle\frac{{p_{{\rm e}\,\alpha}\,n^{\alpha}}}
{{p_{{\rm e}\,\mu}\,U^{\mu}}}\, ,
\end{equation}
where $n^\alpha=-e_{(\theta)}^\alpha$ is the normal to the disc with
respect to the observer co-moving with the matter in the disc.

The relative time delay $\Delta t$ is the Boyer-Lindquist time which elapses
between the emission
of a photon from the disc and its reception by an observer (plus a certain
constant so that the delay is finite close to the black hole). We have integrated
the equation of the geodesics in Kerr ingoing coordinates and thus we have calculated
the delay in the Kerr ingoing time coordinate $\Delta t_{\rm K}$. The Boyer-Lindquist
time coordinate can be obtained from the Kerr ingoing one by the following equation
\begin{equation}
{\rm d}t = {\rm d}t_{\rm K}-\left [ 1+ \frac{2r}{(r-r_+)(r-r_-)}\right ]
{\rm d}r\, ,
\end{equation}
with $r_{\pm}=1\pm\sqrt{1-a^2}$.
Then we define the delay as
\begin{eqnarray}
\label{delay1}
\Delta t & = & \Delta t_{\rm K}-\left [r+\frac{2}{r_+ - r_-}\,
\ln{\frac{r-r_+}{r-r_-}}+\ln{[(r-r_+)(r-r_-)]}\right ]
\quad {\rm for}\quad a < 1\, , \\
\label{delay2}
\Delta t & = & \Delta t_{\rm K}-\left [r-\frac{2}{r-1}+2\ln{(r-1)}\right ]
\quad {\rm for}\quad a = 1\, .
\end{eqnarray}
There is a minus in front of the brackets because the direction of integration
is from infinity (represented by $r_{\rm o}=10^{11}$ in our computations) to the
disc.

The change of the polarization angle is (see \cite{connors_1977},
\cite{connors_1980})
\begin{equation}
\label{polar}
\tan{\Psi}=\frac{Y}{X}\, ,
\end{equation}
where
\begin{eqnarray}
X & = & -(\alpha-a\sin{\theta_{\rm o}})\kappa_1-\beta\kappa_2\, ,\\
Y & = & \phantom{-}(\alpha-a\sin{\theta_{\rm o}})\kappa_2-\beta\kappa_1\, ,\\
\end{eqnarray}
with $\kappa_1$ and $\kappa_2$ being components of the complex constant
of motion $\kappa_{\rm pw}$ (see \cite{walker_1970})
\begin{eqnarray}
\kappa_1 & = & ar\,p_{\rm e}^\theta f^t-r\,[a\,p_{\rm e}^t-(r^2+a^2)\,
p_{\rm e}^\varphi]f^\theta-r(r^2+a^2)\,p_{\rm e}^\theta f^\varphi\, ,\\
\kappa_2 & = & -r\,p_{\rm e}^rf^t+r\,[p_{\rm e}^t-a\,p_{\rm e}^\varphi]f^r+ar\,
p_{\rm e}^rf^\varphi\, .
\end{eqnarray}
Here the polarization vector $f^\mu$ is a four-vector corresponding to the
three-vector $\vec{f_1}$ from Figure~\ref{pol_angle} which is chosen in such
a way that it is a unit vector parallel with $\vec{n'_1}$ (i.e.\ $\Psi_1=0$)
\begin{equation}
f^\mu = \frac{n^\mu-\mu_{\rm e}\left( g\,p_{\rm e}^\mu-U^\mu\right)}
{\sqrt{1-\mu_{\rm e}^2}}\, .
\end{equation}

We define the azimuthal emission angle as the angle between the
projection of the three-momentum of the emitted photon into the disc (in the
local rest frame co-moving with the disc) and the radial tetrad vector:
\begin{equation}
\label{azim_angle}
\Phi_{\rm e}=-{\rm R}_{\rm sgn}^{\rm e}\arccos\left(
\frac{g\,{p_{{\rm e}\,\alpha}}\,e_{(\varphi)}^{\alpha}}
{\sqrt{1-\mu_{\rm e}^2}}\right)+\frac{\pi}{2}\, ,
\end{equation}
where ${\rm R}_{\rm sgn}^{\rm e}$ is positive if the emitted photon travels
outwards ($p_{\rm e}^{(r)}>0$) and negative if it travels inwards
($p_{\rm e}^{(r)}<0$) in the local rest frame of the disc.

We conclude this section by the relationship between the Boyer-Lindquist
coordinate $\varphi$ and the Kerr ingoing coordinate $\varphi_{\rm K}$, which we
use when we interpolate between the pre-calculated tables
\begin{eqnarray}
\varphi & = & \varphi_{\rm K}+\frac{a}{r_+-r_-}\ln{\frac{r-r_+}{r-r_-}}\quad {\rm for}
\quad a<1\, ,\\
\varphi & = & \varphi_{\rm K}-\frac{1}{r-1}\quad {\rm for}\quad a=1\, .
\end{eqnarray}

\subsection{Local emission in lamp-post models}
\label{appendix2}
The local emission from a disc is proportional to the incident illumination from
a power-law primary source placed on the axis at height $h$ above the black
hole. To calculate the incident illumination we need to integrate the geodesics
from the source to the disc.

The four-momentum of the incident photons which were emitted by a primary source and
which are striking the disc at radius $r$ is
(see e.g.\ \cite{carter_1968} and \cite{misner_1973})
\begin{eqnarray}
p_{\rm i}^t & = & 1+2/r+4/\Delta\, ,\\
p_{\rm i}^r & = & {\rm R}^{\prime}_{\rm sgn}[(r^2+a^2)^2-\Delta
(a^2+q_{\rm L}^2)]^{1/2}/r^2\, ,\\
p_{\rm i}^\theta & = & q_{\rm L}/r^2\, ,\\
p_{\rm i}^\varphi & = & 2a/(r\Delta)\, ,
\end{eqnarray}
where
$q_{\rm L}^2 = \sin^2{\!\theta_{\rm L}}\,(h^2+a^2)^2/\Delta_{\rm L}-a^2$ is
Carter's constant of motion with $\Delta_{\rm L}=h^2-2h+a^2$, and with the angle
of emission $\theta_{\rm L}$ being the local angle
under which the photon is emitted from a primary source (it is measured in the
rest frame of the source). We define this angle by
$\tan{\theta_{\rm L}}=-{p_{\rm L}^{(\theta)}/p_{\rm L}^{(r)}}$, where
$p_{\rm L}^{(r)}=p_{\rm L}^\mu\,e_{{\rm L}\,\mu}^{(r)}$ and
$p_{\rm L}^{(\theta)}=p_{\rm L}^\mu\,e_{{\rm L}\,\mu}^{(\theta)}$ with
$p_{\rm L}^\mu$ and $e_{{\rm L}\,\mu}^{(a)}$ being the four-momentum of emitted
photons and the local tetrad connected with a primary source, respectively. The
angle is $0^\circ$ when the photon is emitted downwards and $180^\circ$ if the
photon is emitted upwards.

We denoted the sign of the radial component of the momentum by
${\rm R}^{\prime}_{\rm sgn}$. We have chosen such an affine parameter for the
light geodesic that the conserved energy of the light is
$-p_{{\rm i}\,t}=-p_{{\rm L}\,t}=1$. The conserved angular momentum of incident
photons is zero ($l_{\rm L}=0$).

The gravitational and Doppler shift of the photons striking the disc which were emitted by
a primary source is
\begin{equation}
\label{gfac_lamp}
g_{\rm L}= \frac{\nu_{\rm i}}{\nu_{\rm L}}=
\frac{p_{{\rm i}\,\mu} U^\mu}{p_{{\rm L}\,\alpha} U_{\rm L}^\alpha}=
-\frac{p_{{\rm i}\,\mu} U^\mu}{U_{\rm L}^t}\, .
\end{equation}
Here $\nu_{\rm i}$ and $\nu_{\rm L}$ denote the frequency of the incident and
emitted photons, respectively and $U_{\rm L}^\alpha$ is a four-velocity of the
primary source with the only
non-zero component \hbox{$U_{\rm L}^t=\sqrt{\frac{h^2+a^2}{\Delta_{\rm L}}}$}.


The cosine of the local incident angle is
\begin{equation}
\label{cosine_inc}
\mu_{\rm i}=|\cos{\delta_{\rm i}}\,|=
\displaystyle\frac{{p_{{\rm i}\,\alpha}\,n^{\alpha}}}
{{p_{{\rm i}\,\mu}\,U^{\mu}}}\, ,
\end{equation}
where $n^\alpha=-e_{(\theta)}^\alpha$ is normal to the disc with respect to the
observer co-moving with the matter in the disc.

We further define the azimuthal incident angle as the angle between the
projection of the three-momentum of the incident photon into the disc (in the
local rest frame co-moving with the disc) and the radial tetrad vector,
\begin{equation}
\label{azim_angle_inc}
\Phi_{\rm i}=-{\rm R}_{\rm sgn}^{\rm i}\arccos\left(
\frac{-1}{\sqrt{1-\mu_{\rm i}^2}}\frac{\,{p_{{\rm i}\,\alpha}}
\,e_{(\varphi)}^{\alpha}}
{p_{{\rm i}\,\mu}U^\mu}\right)+\frac{\pi}{2}\, ,
\end{equation}
where ${\rm R}_{\rm sgn}^{\rm i}$ is positive if the incident photon travels
outwards ($p_{\rm i}^{(r)}>0$) and negative if it travels inwards
($p_{\rm i}^{(r)}<0$) in the local rest frame of the disc.

In lamp-post models the emission of the disc will be proportional to the incident
radiation $N_{\rm i}^{S}(E_{\rm l})$ which comes from a primary source
\begin{equation}
\label{illumination}
N_{\rm i}^{S}(E_{\rm l})=N_{\rm L}^{\Omega}(E_{\rm L})\frac{{\rm d}
\Omega_{\rm L}}{{\rm d}S_{\loc}}\, .
\end{equation}
Here $N_{\rm L}^{\Omega}(E_{\rm L})=N_{0{\rm L}}\,E_{\rm L}^{-\Gamma}$ is
an isotropic and stationary power-law emission from a primary source which is
emitted into
a solid angle ${\rm d}\Omega_{\rm L}$ and which illuminates local area
${\rm d}S_{\rm l}$ on the disc. The energy of the photon striking the disc
(measured in the local frame co-moving with the disc) will be redshifted
\begin{equation}
E_\loc=g_{\rm L}\,E_{\rm L}\, .
\end{equation}
The ratio ${\rm d}\Omega_{\rm L}/{\rm d}S_{\loc}$ is
\begin{equation}
\frac{{\rm d}\Omega_{\rm L}}{{\rm d}S_{\loc}}=\frac{{\rm d}\Omega_{\rm L}}
{{\rm d}S} \frac{{\rm d}S}{{\rm d}S_\loc}=\frac{\sin{\theta_{\rm L}{\rm d}
\theta_{\rm L}\,{\rm d}\varphi}}{{\rm d}r\,{\rm d}\varphi}\frac{{\rm d}S}
{{\rm d}S_\loc}\, ,
\end{equation}
where (see eqs.~(\ref{dS2}) and (\ref{dS3}))
\begin{equation}
{\rm d}S={\rm d}r\,{\rm d}\varphi=|{\rm d}^2{\!S_t}^{\theta}|=-g^{\theta\theta}\,
\frac{p_{{\rm i} t}}{p_{\rm i}^\theta}\,{\rm d}S_\perp=
\,\frac{g^{\theta\theta}}{p_{\rm i}^\theta}\,{\rm d}S_\perp\, .
\end{equation}
Here we used
the same space-time slice as in the discussion above eq.~(\ref{dS}) and thus
the element
${\rm d}^2\!S_{\alpha\beta}$ is defined as before, see eq.~(\ref{dS2}). Note
that here the area ${\rm d}S_\perp$ is defined by the incident flux tube as
opposed to ${\rm d}S_\perp$ in eq.~(\ref{ratio_dS}) where it was defined by the
emitted flux tube.
The coordinate area ${\rm d}S$ corresponds to the proper area ${\rm d}S_\perp$
which is perpendicular to the incident light ray (in the local rest frame
co-moving with the disc). The corresponding proper area (measured in the same local
frame) lying in the equatorial plane will be
\begin{eqnarray}
\nonumber
{\rm d}S_{\loc} & = &
|{\rm d}^2{\!S_{(t)}}^{\!\!(\theta)}|=|e_{(t)}^\mu\,e^{{(\theta)}\,\nu}\,
{\rm d}^2\!S_{\mu\nu}|=|g_{\theta\theta}^{-1/2}\,U^{\mu}\,
{\rm d}^2\!S_{\mu\theta}|=\\
& = & -g_{\theta\theta}^{-1/2}\,
\frac{p_{{\rm i} \mu}\,U^\mu}{p_{\rm i}^\theta}\,{\rm d}S_\perp =
g_{\theta\theta}^{-1/2}\frac{U_{\rm L}^t}{p_{\rm i}^\theta}\,g_{\rm L}\,
{\rm d}S_\perp\, .
\label{dSl}
\end{eqnarray}
Here we have used eq.~(\ref{dS}) for the tetrad components of the element
${\rm d}^2\!S_{\alpha\beta}$, eqs.~(\ref{dSperp}) and (\ref{gfac_lamp}).

It follows from eqs.~(\ref{illumination})--(\ref{dSl}) that the incident
radiation will be again a power law with the same photon index
$\Gamma$ as in primary emission
\begin{equation}
N_{\rm i}^{S}(E_{\loc})=N_{0{\rm i}}\,E_{\loc}^{-\Gamma}\, ,
\end{equation}
with the normalization factor
\begin{equation}
N_{0 {\rm i}}=N_{0 {\rm L}}\,g_{\rm L}^{\Gamma-1}\,
\sqrt{1-\frac{2h}{h^2+a^2}}\,\frac{\sin{\theta_{\rm L}}\,
{\rm d}\theta_{\rm L}}{r\,{\rm d}r}\, .
\end{equation}
The emission of the disc due to illumination will be proportional to
this factor.

\subsection{Description of FITS files}
\subsubsection{Transfer functions in
{\fontfamily{pcr}\fontshape{tt}\selectfont KBHtablesNN.fits}}
\label{appendix3a}

The transfer functions are stored in the file {\tt KBHtablesNN.fits} as binary
extensions and parametrized by the value of the observer inclination angle
$\theta_{\rm o}$ and the horizon of the black hole $r_{\rm h}$. We found
parametrization by  $r_{\rm h}$ more convenient than using the rotational
parameter $a$, although the latter choice may be more common. Each
extension provides values of a particular transfer function for different
radii, which are given in terms of $r-r_{\rm h}$, and for the Kerr
ingoing axial coordinates $\varphi_{\rm K}$. Values of the horizon
$r_{\rm h}$, inclination $\theta_{\rm o}$, radius $r-r_{\rm h}$ and
angle $\varphi_{\rm K}$, at which the functions are evaluated, are defined as
vectors at the beginning of the FITS file.

\bigskip
The definition of the file {\tt KBHtablesNN.fits}:
\begin{enumerate} \itemsep -2pt
 \item[0.] All of the extensions defined below are binary.
 \item The first extension contains six integers defining which of the
   functions is present in the tables. The integers correspond to the delay,
   $g$-factor, cosine of the local emission angle, lensing, change of
   the polarization angle
   and azimuthal emission angle, respectively. Value $0$
   means that the function is not present in the tables, value $1$ means it is.
 \item The second extension contains a vector of the horizon values in $GM/c^2$
   ($1.00 \le r_{\rm h} \le 2.00$).
 \item The third extension contains a vector of the values of the observer's
   inclination angle $\theta_{\rm o}$ in degrees
   ($0^\circ \le \theta_{\rm o} \le 90^\circ$, $0^\circ$ -- axis, $90^\circ$
   -- equatorial plane).
 \item The fourth extension contains a vector of the values of the radius
 relative to the  horizon $r-r_{\rm h}$ in $GM/c^2$.
 \item The fifth extension contains a vector of the values of the azimuthal
 angle $\varphi_{\rm K}$ in radians ($0 \le \varphi_{\rm K} \le
    2\pi$). Note that $\varphi_{\rm K}$ is a Kerr ingoing axial coordinate,
    not the Boyer-Lindquist one!
 \item All the previous vectors have to have values sorted in an increasing order.
 \item In the following extensions the transfer functions are defined, each
 extension is for a particular value of $r_{\rm h}$ and $\theta_{\rm o}$.
 The values of $r_{\rm h}$ and $\theta_{\rm o}$ are changing with each
 extension in the following order:\\
    \parbox{\textwidth}{
    {\begin{centering}
    $r_{\rm h}[1] \times \theta_{\rm o}[1]$,\\
    $r_{\rm h}[1] \times \theta_{\rm o}[2]$,\\
    $r_{\rm h}[1] \times \theta_{\rm o}[3]$,\\
    \dots \\
    \dots \\
    $r_{\rm h}[2] \times \theta_{\rm o}[1]$,\\
    $r_{\rm h}[2] \times \theta_{\rm o}[2]$,\\
    $r_{\rm h}[2] \times \theta_{\rm o}[3]$,\\
    \dots \\
    \dots \\
    \end{centering}}}
 \item Each of these extensions has the same number of columns (up to six).
 In each column, a particular transfer function is stored -- the delay,
 $g$-factor, cosine of the local emission angle, lensing, change of the
 polarization  angle and azimuthal emission angle, respectively. The order of
 the functions is important but some of the functions may be missing as defined
 in the first extension (see 1.\ above). The functions are:
   \begin{description} \itemsep -2pt
    \item[\rm delay] -- the Boyer-Lindquist time in $GM/c^3$ that elapses
		between the emission of a photon from the disc and absorption of the photon
		by the observer's eye at infinity plus a constant,
    \item[\rm $g$-factor] -- the ratio of the energy of a photon received by the
    observer at infinity to the local energy of the same photon when emitted
    from an accretion disc,
    \item[\rm cosine of the emission angle] -- the cosine of the local emission
		angle between the emitted light ray and local disc normal,
    \item[\rm lensing] -- the ratio of the area at infinity perpendicular to the
    light rays through which photons come to the proper area on the disc
		perpendicular to the light rays and corresponding to the same flux tube,
    \item[\rm change of the polarization angle in radians] --  if the light
		emitted from the disc is linearly polarized then the direction of
		polarization will be changed by this angle at infinity --
    counter-clockwise if positive, clockwise if negative (we are looking
    towards the coming emitted beam); on the disc we measure the angle of
    polarization with respect to the ``up'' direction perpendicular to the disc
    with respect to the local rest frame; at infinity we also measure the
    angle of polarization with respect to the ``up'' direction perpendicular to
    the disc -- the change of the polarization angle is the difference between
    these two angles,
    \item[\rm azimuthal emission angle in radians] -- the angle between the
    projection of the three-mo\-men\-tum of an emitted photon into the disc (in the
    local rest frame co-moving with the disc) and the radial tetrad vector.
   \end{description}
	 For mathematical formulae defining the
functions see eqs.~(\ref{gfac}), (\ref{lensing})--(\ref{cosine}),
(\ref{delay1})--(\ref{polar}) and (\ref{azim_angle}) in Appendix~\ref{appendix1}.
 \item Each row corresponds to a particular value of $r-r_{\rm h}$
 (see 4.\ above).
 \item Each element corresponding to a particular column and row is a vector.
 Each element of this vector corresponds to a particular value of
 $\varphi_{\rm K}$ (see 5.\ above).
\end{enumerate}

We have pre-calculated three sets of tables -- {\tt KBHtables00.fits},
{\tt KBHtables50.fits} and {\tt KBHtables99.fits}.
All of these tables were computed for an accretion disc near a Kerr black
hole with no disc corona present. Therefore, ray-tracing in the vacuum Kerr
space-time could be used for calculating the transfer functions.
When computing the transfer functions, it was supposed that the matter in
the disc rotates on stable circular (free) orbits above the marginally stable
orbit. The matter below this orbit is freely falling and has the same energy and
angular momentum as the matter which is on the marginally stable orbit.

The observer is placed in the direction $\varphi = \pi/2$. The black hole
rotates counter-clockwise. All six functions are present in these tables.

Tables are calculated for these values of the black-hole horizon:\\
-- {\tt KBHtables00.fits}: 1.00, 1.05, 1.10, 1.15, \dots, 1.90, 1.95, 2.00
(21 elements),\\
-- {\tt KBHtables50.fits}: 1.00, 1.10, 1.20, \dots, 1.90, 2.00 (11 elements),\\
-- {\tt KBHtables99.fits}: 1.05 (1 element),\\
and for these values of the observer's inclination:\\
-- {\tt KBHtables00.fits}: 0.1, 1, 5, 10, 15, 20, \dots, 80, 85, 89
(20 elements),\\
-- {\tt KBHtables50.fits}: 0.1, 1, 10, 20, \dots, 80, 89 (11 elements),\\
-- {\tt KBHtables99.fits}: 0.1, 1, 5, 10, 15, 20, \dots, 80, 85, 89
(20 elements).

The radii and azimuths at which the functions are evaluated are same for all
three tables:\\
-- radii $r-r_{\rm h}$ are exponentially increasing from 0 to 999
(150 elements),\\
-- values of the azimuthal angle $\varphi_{\rm K}$ are equidistantly spread from
0 to $2\pi$ radians with a much denser cover ``behind'' the black hole, i.e.\
near $\varphi_{\rm K} = 1.5\pi$
(because some of the functions are changing heavily in this area for higher
inclination angles, $\theta_{\rm o} > 70^\circ$) (200 elements).

\subsubsection{Tables in
{\fontfamily{pcr}\fontshape{tt}\selectfont KBHlineNN.fits}}
\label{appendix3b}
Pre-calculated functions ${\rm d}F(g)\equiv{\rm d}g\,F(g)$ defined in
eq.~(\ref{conv_function}) are stored in FITS files
{\tt KBHlineNN.fits}. These functions are used by all axisymmetric models.
They are stored as binary extensions and they are parametrized by the value of
the observer inclination angle $\theta_{\rm o}$ and the horizon of the black
hole $r_{\rm h}$. Each extension provides values for different
radii, which are given in terms of $r-r_{\rm h}$, and for different
$g$-factors. Values of the $g$-factor, radius $r-r_{\rm h}$, horizon
$r_{\rm h}$, and inclination $\theta_{\rm o}$, at which the functions are
evaluated, are defined as vectors at the beginning of the FITS file.

The definition of the file {\tt KBHlineNN.fits}:
\begin{enumerate} \itemsep -2pt
 \item[0.] All of the extensions defined below are binary.
 \item The first extension contains one row with three columns that define
 bins in the $g$-factor:
  \begin{list}{--}{\setlength{\topsep}{-2pt}\setlength{\itemsep}{-2pt}
	                 \setlength{\leftmargin}{1em}}
	 \item integer in the first column defines the width of the bins
	 (0 -- constant, 1 -- exponentially growing),
   \item real number in the second column defines the lower boundary of the
	 first bin (minimum of the $g$-factor),
   \item real number in the third column defines the upper boundary of the
	 last bin (maximum of the $g$-factor).
	\end{list}
 \item The second extension a contains vector of the values of the radius
 relative to the horizon $r-r_{\rm h}$ in $GM/c^2$.
 \item The third extension contains a vector of the horizon values in $GM/c^2$
   ($1.00 \le r_{\rm h} \le 2.00$).
 \item The fourth extension contains a vector of the values of the observer's
   inclination angle $\theta_{\rm o}$ in degrees
   ($0^\circ \le \theta_{\rm o} \le 90^\circ$, $0^\circ$ -- axis, $90^\circ$
   -- equatorial plane).
 \item All the previous vectors have to have values sorted in an increasing
 order.
 \item In the following extensions the functions ${\rm d}F(g)$ are defined, each
 extension is for a particular value of $r_{\rm h}$ and $\theta_{\rm o}$.
 The values of $r_{\rm h}$ and $\theta_{\rm o}$ are changing with each
 extension in the same order as in tables in the {\tt KBHtablesNN.fits} file
 (see the previous section, point 7.). Each extension has one column.
 \item Each row corresponds to a particular value of $r-r_{\rm h}$
 (see 2.\ above).
 \item Each element corresponding to a particular column and row is a vector.
 Each element of this vector corresponds to a value of the function
 for a particular bin in the $g$-factor.
 This bin can be calculated from number of elements of the vector and data
 from the first extension (see 1.\ above).
\end{enumerate}

We have pre-calculated several sets of tables for different limb
darkening/brightening laws and with different resolutions. All of them were
calculated from tables in the {\tt KBHtables00.fits} (see the previous section
for details) and therefore these tables are calculated for the same values of
the black-hole horizon and observer's inclination. All of these tables have
equidistant bins in the $g$-factor which fall in the interval
$\langle0.001,1.7 \rangle$. Several sets of tables are available:\\
-- {\tt KBHline00.fits} for isotropic emission, see eq.~(\ref{isotropic}),\\
-- {\tt KBHline01.fits} for Laor's limb darkening, see eq.~(\ref{laor}),\\
-- {\tt KBHline02.fits} for Haardt's limb brightening, see eq.~(\ref{haardt}).\\
All of these tables have 300 bins in the $g$-factor and 500 values of the radius
$r-r_{\rm h}$ which are exponentially increasing from 0 to 999.
We have produced also tables
with a lower resolution -- {\tt KBHline50.fits}, {\tt KBHline51.fits} and
{\tt KBHline52.fits} with 200 bins in the $g$-factor and 300 values of the
radius.

\subsubsection{Lamp-post tables in
{\fontfamily{pcr}\fontshape{tt}\selectfont lamp.fits}}
\label{appendix3c}

This file contains pre-calculated values of the functions needed for the lamp-post
model. It is supposed that a primary source of emission is placed on the axis at a
height $h$ above
the Kerr black hole. The matter in the disc rotates on stable circular (free) orbits
above the marginally stable orbit and it is freely falling below this orbit where it
has the same energy and angular momentum as the matter which is on the marginally stable
orbit. It is assumed that the corona between the source and the disc is optically thin,
therefore ray-tracing in the
vacuum Kerr space-time could be used for computing the functions.

There are five functions stored in the {\tt lamp.fits} file as binary
extensions. They are parametrized by the value of the horizon of the black hole
$r_{\rm h}$, and height $h$, which are defined  as vectors at the beginning of the
FITS file. Currently only tables for $r_{\rm h}=1.05$ (i.e.\ $a\doteq0.9987492$)
and $h=2,\,3,\,4,\,5,\,6,\,8,\,10,\,12,\,15,\,20,\,30,\,50,\,75$ and $100$ are
available.
The functions included are:
 \begin{list}{}{\setlength{\topsep}{-2pt}\setlength{\itemsep}{-2pt}}
  \item[\rm -- angle of emission in degrees] -- the angle under which a photon is
  emitted from a primary source placed at a height $h$ on the axis above the
  black hole measured by a local stationary observer ($0^\circ$ -- a photon is
	emitted downwards, $180^\circ$
	-- a photon is emitted upwards),
  \item[\rm -- radius] -- the radius in $GM/c^2$ at which a photon strikes the disc,
  \item[\rm -- $g$-factor] -- the ratio of the energy of a photon hitting the disc
  to the energy of the same photon when emitted from a primary source,
  \item[\rm -- cosine of the incident angle] -- an absolute value of the cosine of
  the local incident angle between the incident light ray and local disc normal,
  \item[\rm -- azimuthal incident angle in radians] -- the angle between the
  projection of the three-momentum of the incident photon into the disc (in the
  local rest frame co-moving with the disc) and the radial tetrad vector.
 \end{list}
For mathematical formulae defining the functions see
eqs.~(\ref{gfac_lamp})--(\ref{azim_angle_inc}) in Appendix~\ref{appendix2}.

The definition of the file {\tt lamp.fits}:
\begin{enumerate} \itemsep -2pt
 \item[0.] All of the extensions defined below are binary.
 \item The first extension contains a vector of the horizon values in $GM/c^2$,
   though currently only FITS files with tables for one value of the black-hole
	 horizon are accepted ($1.00 \le r_{\rm h} \le 2.00$).
 \item The second extension contains a vector of the values of heights $h$
 of a primary source in $GM/c^2$.
 \item In the following extensions the functions are defined. Each extension is
 for a particular value of $r_{\rm h}$ and $h$. The values of $r_{\rm h}$ and $h$
 are changing with each extension in the following order:\\
    \parbox{\textwidth}{
    {\begin{centering}
    $r_{\rm h}[1] \times h[1]$,\\
    $r_{\rm h}[1] \times h[2]$,\\
    $r_{\rm h}[1] \times h[3]$,\\
    \dots \\
    \dots \\
    $r_{\rm h}[2] \times h[1]$,\\
    $r_{\rm h}[2] \times h[2]$,\\
    $r_{\rm h}[2] \times h[3]$,\\
    \dots \\
    \dots \\
    \end{centering}}}
 \item Each of these extensions has five columns.
 In each column, a particular function is stored -- the angle of emission, radius,
 $g$-factor, cosine of the local incident angle and azimuthal incident angle,
 respectively. Extensions may have a different number of rows.
\end{enumerate}

\subsubsection{Coefficient of reflection in
{\fontfamily{pcr}\fontshape{tt}\selectfont
fluorescent\_line.fits}}
\label{appendix3d}
Values of the coefficient of reflection $f(\mu_{\rm i},\mu_{\rm e})$ for a
fluorescent
line are stored for different incident and reflection angles in this file.
For details on the model of scattering used for computations see
\cite{matt_1991}. It is assumed that the incident radiation is a power law
with the photon index $\Gamma=1.7$.
The coefficient does not change its angular dependences for other photon indices,
only its normalization changes (see Figure~14 in \cite{george_1991}).
The FITS file consists of three binary extensions:
\begin{list}{}{\setlength{\topsep}{-2pt}\setlength{\itemsep}{-2pt}}
 \item[--] the first extension contains absolute values of the cosine of incident
 angles,
 \item[--] the second extension contains values of the cosine of reflection angles,
 \item[--] the third extension contains one column with vector elements, here
 values of the coefficient of reflection are stored for different incident angles (rows)
 and for different reflection angles (elements of a vector).
\end{list}

\subsubsection{Tables in
{\fontfamily{pcr}\fontshape{tt}\selectfont refspectra.fits}}
\label{appendix3e}
The function $f(E_\loc;\mu_{\rm i},\mu_{\rm e})$ which gives dependence of a
locally emitted spectrum on the angle of incidence and angle of emission
is stored in this FITS file. The emission is induced by a power-law incident
radiation. Values of this function were
computed by the Monte Carlo simulations of Compton scattering, for details see
\cite{matt_1991}. The reflected radiation depends on the photon index $\Gamma$
of the incident radiation. There are several binary extensions in this FITS file:
\begin{list}{}{\setlength{\topsep}{-2pt}\setlength{\itemsep}{-2pt}}
 \item[--] the first extension contains energy values in keV where
 the function $f(E_\loc;\mu_{\rm i},\mu_{\rm e})$ is computed, currently the
 interval from 2 to 300~keV is covered,
 \item[--] the second extension contains the absolute values of the cosine of
 the incident angles,
 \item[--] the third extension contains the values of the cosine of the emission
 angles,
 \item[--] the fourth extension contains the values of the photon indices
 $\Gamma$ of the incident power law, currently tables for
 $\Gamma=1.5,\,1.6,\,\dots,\,2.9$ and  $3.0$ are computed,
 \item[--] in the following extensions the function
 $f(E_\loc;\mu{\rm i},\mu_{\rm e})$ is defined, each extension is
 for a particular value of $\Gamma$; here values of the function are stored as a
 vector for different incident angles (rows) and for different angles of emission
 (columns), each element of this vector corresponds to a value of the function
 for a certain value of energy.
\end{list}

\subsection{Description of the integration routines}
Here we describe the technical details about the integration routines,
which act as a common driver performing the ray-tracing for various
models of the local emission. The description of non-axisymmetric and
axisymmetric versions are both provided.
An appropriate choice depends on the form of intrinsic emissivity.
Obviously, non-axisymmetric tasks are computationally more demanding.

\subsubsection{Non-axisymmetric integration routine
{\fontfamily{pcr}\fontshape{tt}\selectfont ide}}
\label{appendix4a}
This subroutine integrates the local emission and local Stokes parameters
for (partially) polarized emission of the accretion disc near a rotating
(Kerr) black hole (characterized by the angular momentum $a$) for an observer
with an inclination angle $\theta_{\rm o}$.
The subroutine has to be called with ten parameters:
\begin{center}
 {\tt ide(ear,ne,nt,far,qar,uar,var,ide\_param,emissivity,ne\_loc)}
\end{center}
\begin{description} \itemsep -2pt
 \item[{\tt ear}] -- real array of energy bins (same as {\tt ear} for local
 models in {\sc{}xspec}),
 \item[{\tt ne}] -- integer, number of energy bins (same as {\tt ne} for local
 models in {\sc{}xspec}),
 \item[{\tt nt}] -- integer, number of grid points in time (${\tt nt}=1$ means
 stationary model),
 \item[{\tt far(ne,nt)}] -- real array of photon flux per bin
 (same as {\tt photar} for local models in {\sc{}xspec} but with the time
 resolution),
 \item[{\tt qar(ne,nt)}] -- real array of the Stokes parameter Q divided by the
 energy,
 \item[{\tt uar(ne,nt)}] -- real array of the Stokes parameter U divided by the
 energy,
 \item[{\tt var(ne,nt)}] -- real array of the Stokes parameter V divided by the
 energy,
 \item[{\tt ide\_param}] -- twenty more parameters needed for the integration
 (explained below),
 \item[{\tt emissivity}] -- name of the external emissivity subroutine, where
 the local emission of the disc is defined (explained in detail below),
 \item[{\tt ne\_loc}] -- number of points (in energies) where local photon flux
         (per keV) in the emissivity subroutine is defined.
\end{description}

The description of the {\tt ide\_param} parameters follows:
\begin{description} \itemsep -2pt
 \item[{\tt ide\_param(1)}] -- {\tt a/M} -- the black-hole angular momentum
 ($0 \le {\tt a/M} \le 1$),
 \item[{\tt ide\_param(2)}] -- {\tt theta\_o} -- the observer inclination in
degrees ($0^\circ$ -- pole, $90^\circ$ -- equatorial plane),
 \item[{\tt ide\_param(3)}] -- {\tt rin-rh} -- the inner edge of the non-zero
 disc emissivity relative to the black-hole horizon (in $GM/c^2$),
 \item[{\tt ide\_param(4)}] -- {\tt ms} -- determines whether we also integrate
 emission below the marginally stable orbit; if its value is set to zero and
 the inner radius of the disc is below the marginally stable orbit then the
 emission below this orbit is taken into account, if set to unity it is not,
 \item[{\tt ide\_param(5)}] -- {\tt rout-rh} -- the outer edge of the non-zero
 disc emissivity relative to the black-hole horizon (in $GM/c^2$),
 \item[{\tt ide\_param(6)}] -- {\tt phi} -- the position angle of the axial sector
 of the disc with non-zero emissivity in degrees,
 \item[{\tt ide\_param(7)}] -- {\tt dphi} -- the inner angle of the axial sector
 of the disc with non-zero emissivity in degrees
 (${\tt dphi} \le 360^\circ$),
 \item[{\tt ide\_param(8)}] -- {\tt nrad} -- the radial resolution of the grid,
 \item[{\tt ide\_param(9)}] -- {\tt division} -- the switch for the spacing of
 the radial grid ($0$ -- equidistant, $1$ -- exponential),
 \item[{\tt ide\_param(10)}] -- {\tt nphi} -- the axial resolution of the grid,
 \item[{\tt ide\_param(11)}] -- {\tt smooth} -- the switch for performing simple
 smoothing ($0$ -- no, $1$ -- yes),
 \item[{\tt ide\_param(12)}] -- {\tt normal} -- the switch for normalizing the
 final spectrum,\\
   if $=$ 0 -- total flux is unity (usually used for the line),\\
   if $>$ 0 -- flux is unity at the energy = {\tt normal} keV (usually used for
   the continuum),\\
   if $<$ 0 -- final spectrum is not normalized,
 \item[{\tt ide\_param(13)}] -- {\tt zshift} -- the overall redshift of the
 object,
 \item[{\tt ide\_param(14)}] -- {\tt ntable} -- tables to be used, it defines
  a double digit number NN in the name of the FITS file KBHtablesNN.fits
	containing the tables ($0 \le {\tt ntable} \le 99$),
 \item[{\tt ide\_param(15)}] -- {\tt edivision} -- the switch for spacing the
 grid in local energies (0 -- equidistant, 1 -- exponential),
 \item[{\tt ide\_param(16)}] -- {\tt periodic} -- if set to unity then local
 emissivity is periodic if set to zero it is not (need not be set if
 ${\tt nt}=1$),
 \item[{\tt ide\_param(17)}] -- {\tt dt} -- the time step (need not be set if
 ${\tt nt}=1$),
 \item[{\tt ide\_param(18)}] -- {\tt polar} -- whether the change of the
 polarization angle and/or azimuthal emission angle will be read from FITS
 tables  (0 -- no, 1 -- yes),
 \item[{\tt ide\_param(19)}] -- {\tt r0-rh} and
 \item[{\tt ide\_param(20)}] -- {\tt phi0} -- in dynamical computations the
 initial time will be set to the time when photons emitted from the point
 [{\tt r0}, {\tt phi0}] on the disc (in the Boyer-Lindquist coordinates) reach
 the observer.
\end{description}

The {\tt ide} subroutine needs an external emissivity subroutine in which the
local emission and local Stokes parameters are defined. This subroutine has
twelve parameters:\\[3mm]
\hspace*{5mm}{\tt emissivity(\parbox[t]{\textwidth}
{ear\_loc,ne\_loc,nt,far\_loc,qar\_loc,uar\_loc,var\_loc,r,phi,cosine,\\
 phiphoton,first\_emis)}}
\begin{description} \itemsep -2pt
 \item[{\tt ear\_loc(0:ne\_loc)}] -- real array of the local energies where
 local photon flux {\tt far\_loc} is defined, with special meaning of
 {\tt ear\_loc(0)} -- if its value is larger than zero then the local emissivity
 consists of two energy regions where the flux is non-zero; the flux between
 these regions is zero and {\tt ear\_loc(0)} defines the number of points in
 local energies with the zero local flux,
 \item[{\tt ne\_loc}] -- integer, the number of points (in energies) where the
 local photon flux (per keV) is defined,
 \item[{\tt nt}] -- integer, the number of grid points in time (${\tt nt}=1$
 means stationary model),
 \item[{\tt far\_loc(0:ne\_loc,nt)}] -- real array of the local photon flux
 (per keV) -- if the local emissivity consists of two separate non-zero regions
    (i.e.\ ${\tt ear\_loc(0)} > 0$) then {\tt far\_loc(0,it)} is the index of
    the last point of the first non-zero local energy region,
 \item[{\tt qar\_loc(ne\_loc,nt)}] -- real array of the local Stokes parameter Q
 divided by the local energy,
 \item[{\tt uar\_loc(ne\_loc,nt)}] -- real array of the local Stokes parameter U
 divided by the local energy,
 \item[{\tt var\_loc(ne\_loc,nt)}] -- real array of the local Stokes parameter V
 divided by the local energy,
 \item[{\tt r}] -- the radius in $GM/c^2$ where the local photon flux
 {\tt far\_loc} at the local energies {\tt ear\_loc} is demanded
 \item[{\tt phi}] -- the azimuth (the Boyer-Lindquist coordinate $\varphi$)
 where the local photon flux {\tt far\_loc} at the local energies {\tt ear\_loc}
 is demanded,
 \item[{\tt cosine}] -- the cosine of the local angle between the emitted ray and
 local disc normal,
 \item[{\tt phiphoton}] -- the angle between the emitted ray projected onto the
 plane of the disc (in the local frame of the moving disc) and the radial
 component of the local tetrad (in radians),
 \item[{\tt first\_emis}] -- boolean, TRUE if we enter the emissivity subroutine
 from the {\tt ide} subroutine for the first time, FALSE if this subroutine was
 already evaluated during the present run. This distinction is convenient 
 to initialize some variables when calling the emissivity subroutine for the 
 first time (e.g.\ calculation of the falling spot trajectory can be performed
 in this place).
\end{description}

\subsubsection{Axisymmetric integration routine
{\fontfamily{pcr}\fontshape{tt}\selectfont idre}}
\label{appendix4b}
This subroutine integrates the local axisymmetric emission of an accretion disc
near a rotating (Kerr) black hole (characterized by the angular momentum $a$)
for an observer with an inclination angle~$\theta_{\rm o}$.
The subroutine has to be called with eight parameters:
\begin{center}
 {\tt idre(ear,ne,photar,idre\_param,cmodel,ne\_loc,ear\_loc,far\_loc)}
\end{center}
\begin{description} \itemsep -2pt
 \item[{\tt ear}] -- real array of energy bins (same as {\tt ear} for local
 models in {\sc{}xspec}),
 \item[{\tt ne}] -- integer, the number of energy bins (same as {\tt ne} for local
 models in {\sc{}xspec}),
 \item[{\tt photar}] -- real array of the photon flux per bin
 (same as {\tt photar} for local models in {\sc{}xspec}),
 \item[{\tt idre\_param}] -- ten more parameters needed for the integration
 (explained below),
 \item[{\tt cmodel}] -- 32-byte string with a base name of a FITS file with
 tables for axisymmetric emission (e.g.\ ``{\tt KBHline}'' for
 {\tt KBHlineNN.fits}),
 \item[{\tt ne\_loc}] -- the number of points (in energies) where the local
 photon flux (per keV) is defined in the emissivity subroutine,
 \item[{\tt ear\_loc}] -- array of the local energies where the local
 photon flux {\tt far\_loc} is defined
 \item[{\tt far\_loc}] -- array of the local photon flux (per keV).
\end{description}

The description of the {\tt idre\_param} parameters follows:
\begin{description} \itemsep -2pt
 \item[{\tt idre\_param(1)}] -- {\tt a/M} -- the black-hole angular momentum
 ($0 \le {\tt a/M} \le 1$),
 \item[{\tt idre\_param(2)}] -- {\tt theta\_o} -- the observer inclination in
degrees ($0^\circ$ -- pole, $90^\circ$ -- equatorial plane),
 \item[{\tt idre\_param(3)}] -- {\tt rin-rh} -- the inner edge of the non-zero
 disc emissivity relative to the black-hole horizon (in $GM/c^2$),
 \item[{\tt idre\_param(4)}] -- {\tt ms} -- determines whether we also integrate
emission below the marginally stable orbit; if its value is set to zero and
the inner radius of the disc is below the marginally stable orbit then the
emission below this orbit is taken into account, if set to unity it is not,
 \item[{\tt idre\_param(5)}] -- {\tt rout-rh} -- the outer edge of the non-zero
 disc emissivity relative to the black-hole horizon (in $GM/c^2$),
 \item[{\tt idre\_param(6)}] -- {\tt smooth} -- the switch for performing simple
 smoothing ($0$ -- no, $1$ -- yes),
 \item[{\tt idre\_param(7)}] -- {\tt normal} -- the switch for normalizing the
 final spectrum,\\
   if $=$ 0 -- total flux is unity (usually used for the line),\\
   if $>$ 0 -- flux is unity at the energy = {\tt normal} keV (usually used for
   the continuum),\\
   if $<$ 0 -- final spectrum is not normalized,
 \item[{\tt idre\_param(8)}] -- {\tt zshift} -- the overall redshift of the object,
 \item[{\tt idre\_param(9)}] -- {\tt ntable} -- tables to be used, it defines
 a double digit number NN in the name of the FITS file (e.g.\ in
 {\tt KBHlineNN.fits}) containing the tables ($0 \le {\tt ntable} \le 99$),
 \item[{\tt idre\_param(10)}] -- {\tt alpha} -- the radial power-law index.
\end{description}

The {\tt idre} subroutine does not need for its operation any external 
emissivity subroutine.


\begin{thebibliography}{99}
\footnotesize
\parskip-2pt
\bibitem{arnaud96}
 Arnaud K.~A., 1996. XSPEC: The first ten years. In {\it Astronomical Data
 Analysis Software and
 Systems V}, eds.\ Jacoby~G. \& Barnes~J., ASP Conf.\ Series, vol.~101,
 p.~17
\bibitem{beckwith04}
 Beckwith~K., \& Done~C., 2004. Iron line profiles in strong gravity.
 {\em MNRAS}, in press (astro-ph/0402199)
\bibitem{carter_1968}Carter B., 1968. Global structure of the Kerr family of 
 gravitational fields. {\em Phys. Rev.}, 174, 1559
\bibitem{chandrasekhar_1960}
 Chandrasekhar S.\ 1960. {\it Radiative Transfer}. Dover publications, New York
\bibitem{chandra92}
 Chandrasekhar S., 1992. {\it The Mathematical Theory of Black Holes.}
 New York, Oxford University Press
\bibitem{connors_1977}Connors P.~A., \& Stark R.~F., 1977. Observable gravitational
 effects on polarized radiation coming from near a black hole. {\em Nature}, 269, 128
\bibitem{connors_1980}Connors P.~A., Piran T., \& Stark R.~F., 1980. Polarization
 features of X-ray radiation emitted near black holes. {\em ApJ}, 235, 224
\bibitem{dovciak04a}
 Dov\v{c}iak M., 2004. PhD Thesis (Charles University, Prague), in preparation
\bibitem{dovciak04}
 Dov\v{c}iak M., Karas V., \& Yaqoob T., 2004. An extended scheme for fitting X-ray
 data with accretion disc spectra in the strong gravity regime. {\em ApJS},
 153, 205 
\bibitem{fabian00} Fabian A.~C., Iwasawa K., Reynolds C.~S., \& Young A.~J., 2000.
 Broad iron lines in active galactic nuclei. {\em PASP}, 112, 1145
\bibitem{fanton97} Fanton~C., Calvani~M., de Felice~F., \& \v{C}ade\v{z}~A.,
 1997. Detecting accretion discs in active galactic nuclei. {\em PASJ}, 49, 159
\bibitem{george_1991} George I.~M., \& Fabian A.~C., 1991. X-ray reflection from
 cold matter in active galactic nuclei and X-ray binaries. {\em MNRAS}, 249, 352
\bibitem{ghisellini_1994} Ghisellini G., Haardt F., \& Matt~G., 1994.
 The contribution of the obscuring torus to the X-ray spectrum of Seyfert
 galaxies -- a test for the unification model. {\em MNRAS}, 267, 743
\bibitem{gierlinski01} Gierli\'{n}ski M., Maciolek-Nied\'{z}wiecki~A.,
 \& Ebisawa~K., 2001. Application of a relativistic accretion disc model to X-ray
 spectra of LMC X-1 and GRO J1655-40. {\em MNRAS}, 325, 1253
\bibitem{haardt_1993} Haardt F.\ 1993. Anisotropic Comptonization in thermal
 plasmas -- Spectral distribution in plane-parallel geometry. {\em ApJ}, 413, 680
\bibitem{kato98} Kato S., Fukue J., \& Mineshige S., 1998.
 {\it Black-Hole Accretion Discs.} Kyoto, Kyoto Univ.\ Press
\bibitem{krolik99} Krolik J.~H., 1999. {\it Active Galactic Nuclei.}
 Princeton University Press, Princeton
\bibitem{laor_1990} Laor A., Netzer H., \& Piran, T., 1990. Massive thin
 accretion discs. II -- Polarization. {\em MNRAS}, 242, 560
\bibitem{laor_1991} Laor A.\ 1991. Line profiles from a disc around
 a rotating black hole. {\em ApJ}, 376, 90
\bibitem{martocchia00} Martocchia A., Karas V., \& Matt G., 2000. Effects of Kerr
 space-time on spectral features from X-ray illuminated accretion discs.
 {\em MNRAS}, 312, 817
\bibitem{matt_1991} Matt G., Perola G.~C., \& Piro L., 1991. The iron line
 and high energy bump as X-ray signatures of cold matter in Seyfert 1 galaxies.
{\em A\&A}, 247, 25
\bibitem{matt92} Matt G., Perola G.~C., Piro L., \& Stella~L., 1992. Iron K-alpha
 line from X-ray illuminated relativistic discs. {\em A\&A}, 257, 63;
 {ibid.} 1992, 263, 453
\bibitem{misner_1973} Misner C.~W., Thorne K.~S., \& Wheeler J.~A., 1973.
{\em Gravitation}. San Fransisco, W.H.Freedman \& Co.
\bibitem{nt73} Novikov I.~D., \& Thorne K.~S., 1973. In {\it Black Holes}, eds.\
 DeWitt~C., DeWitt B.~S.\ New York, Gordon \& Breach, p.~343
\bibitem{phillips_1986} Phillips K.~C., \& M\'{e}sz\'{a}ros~P., 1986.
 Polarization and beaming of accretion disc radiation. {\em ApJ}, 310, 284
\bibitem{rauch94} Rauch K.~P., \& Blandford R.~D., 1994. Optical caustics in a Kerr
 space-time and the origin of rapid X-ray variability in active galactic nuclei.
 {\em ApJ}, 421, 46
\bibitem{reynolds03} Reynolds C.~S., \& Nowak M.~A., 2003. Fluorescent iron lines as
 a probe of astrophysical black hole systems. {\em Phys. Rep.}, 377, 389
\bibitem{schnittman04} Schnittman J.~D., \& Bertschinger E., 2004. The harmonic 
 structure of high-frequency quasi-periodic oscillations in accreting black holes. 
 {\em ApJ}, 606, 1098
\bibitem{walker_1970} Walker M., \& Penrose~R., 1970. On quadratic first integrals of
 the geodesic equations for type \{22\} spacetimes. {\em Commun. Math. Phys.},
 18, 265
\end{thebibliography}
\end{document}